\documentclass[aps,prd,superscriptaddress,12pt,showpacs,notitlepage]{revtex4}
    \usepackage{graphicx}
    \usepackage{amsmath,amssymb,mathrsfs}
\usepackage{epsf}
\usepackage{epsfig}
\usepackage[font=small,skip=0pt,justification=raggedright]{caption}

\newcounter{fig}   \newcommand{\lbfig}[1]{\refstepcounter{fig}
\label{#1} }

\newcommand{\Tr}{{\rm Tr}}

\newcommand{\bea}{\begin{eqnarray}}
\newcommand{\eea}{\end{eqnarray}}
\newcommand{\be}{\begin{equation}}
\newcommand{\ee}{\end{equation}}

\newcommand{\re}[1]{(\ref{#1})}

\newcommand{\bb}{\bibitem}

\newcommand{\eqn}{\begin{eqnarray}}
\newcommand{\eqnx}{\end{eqnarray}}

\begin{document}

\title{Generalized Skyrmions and hairy black holes in asymptotically AdS$_4$ spacetime}

\author{I. Perapechka}
\affiliation{Department of Theoretical Physics and Astrophysics, BSU, Minsk, Belarus}
\author{Ya. Shnir}
\affiliation{Department of Theoretical Physics and Astrophysics, BSU, Minsk, Belarus\\
BLTP, JINR, Dubna, Russia}

\begin{abstract}
We investigate the properties of spherically symmetric
black hole solutions in the generalized Einstein-Skyrme model
theory in four-dimensional asymptotically anti-de Sitter space-time.
The dependencies of the Skyrmion fields
on the cosmological constant and on the strength of the effective gravitational coupling are examined.
We show that the increase of  the absolute value of the cosmological constant qualitatively yields the same effect as increasing
of the effective gravitational coupling. We confirm that, similar to the model in the asymptotically flat space-time, a necessary
condition for the existence of black holes with Skyrmionic hair is the inclusion of the Skyrme term.
\end{abstract}
\maketitle

%%%%%%%%%%%%%%%%%%%%%%%%%%%%%%%%%%%%%%%%%
\section{Introduction}
%%%%%%%%%%%%%%%%%%%%%%%%%%%%%%%%%%%%%%%%%
Existence of black holes appear to be an unavoidable consequence of General
Relativity as well as its various extensions. Since ground-breaking work by
Hawking \cite{Hawking:1974sw}, there has been an increasing interest in investigation of
various aspects of these solutions, especially the thermodynamics of the black holes.

When gravity is coupled to the matter field, the complete system of the Einstein equations may support
both the existence of the gravitating spacially localised particle-like solutions and black holes. The
soliton solutions may be linked to the black holes with hair, which represent
gravitating field configurations possessing both
an event horizon and a non trivial structure of the matter fields (see, e.g. \cite{Volkov:1998cc}).

The first example of the hairy black hole solutions was constructed in the
asymptotically flat Einstein-Skyrme model \cite{Luckock:1986tr,Droz:1991cx}. These
solutions are stable, asymptotically flat and possess a regular horizon, furthermore, they may
be viewed as bound states of Skyrmions and Schwarzschild black holes \cite{Ashtekar}.

The corresponding globally regular gravitating Skyrmions in asymptotically flat space
were considered in  many works, in particular it was shown  \cite{Droz:1991cx,Bizon:1992gb}
that there are two branches of solutions. One of these braches is linked to the usual
flat space  Skyrmion solution, while another, higher in energy branch, in the limit of
vanishing effective coupling constant approaches the Bartnik-McKinnon solution of the
$SU(2)$ Einstein-Yang-Mills theory \cite{Bartnik:1988am}. The stability analysis shows \cite{Bizon:1992gb,Maeda:1993ap}
that the lowest in energy branch is stable with respect to linear perturbations, while the upper branch
is instable, it may collapse  to the Schwarzschild black hole.

Various examples of the gravitating Skyrmions and hairy black holes were investigated over last 20 years.
Axially symmetric globally regular Skyrmions and black holes with hair were constructed in \cite{Sawado:2004yq},
self-gravitating multi-Skyrmion configurations with
discrete symmetry were investigated in \cite{Ioannidou:2006mg}, spinning gravitating Skyrmions were studied in
\cite{Ioannidou:2006nn}. Recently self-gravitating axially symmetric sphaleron solutions of the Einstein-Skyrme model
were constructed \cite{Shnir:2015aba}.

An interesting development is related with investigation of the black holes with Skyrmion hair in
anti-de Sitter (AdS) spacetime, in particular
in the context of the AdS/CFT conjecture \cite{Maldacena:1997re,Witten:1998qj}. On the other hand, the Skyrmion-like
solutions arise as configurations, which are
holographically dual to a D4-brane wrapped on the non-trivial four-cycle in the
(4+1)-dimensional bulk spacetime with AdS geometry and conformal boundary \cite{Sakai:2004cn}.

It is known that the hairy black holes in the asymptotically AdS spacetime can be stabilized as the
cosmological constant $\Lambda$ grows \cite{Winstanley:1998sn}.
Indeed, the no-hair conjecture was revisited for the scalar theory in asymptotically AdS spacetime \cite{Torii:2001pg},
it was
confirmed that the spherically symmetric stable black hole solutions with scalar hair can exist in this model.
Further, the Einstein-Skyrme  model in the asymptotically AdS spacetime was considered in \cite{Shiiki:2005aq}, it
was shown that the stable black holes with Skyrmion hair exist up for some maximum value of $|\Lambda|$.
Similar to the case of the solitons of the Einstein-Skyrme model in asymptotically flat spacetime, there are
two branches of solutions for each value of $\Lambda$ as well as the effective gravitational coupling constant.

Over the last decade a number of modifications of the Skyrme model have been
proposed. The main motivation was to improve phenomenological predictions of the
Skyrme model \cite{skyrme}, which was originally suggested as a candidate for low-energy QCD effective theory.
In this framework
the nuclei are considered as topological solitons. The Skyrme Lagrangian may be recovered from the
$1/N$ expansion in derivatives of non-perturbative low-energy QCD effective action, then one has to take into account
the first two terms only. However, this restriction is not very natural, thus
it was suggested to generalize the
original Skyrme model by inclusion into Lagrangian some additional terms which
are higher order either in derivatives \cite{Marleau:1991jk,Neto:1994bu,Floratos:2001ih,Adam:2010fg}.
The natural extension of  the basic Skyrme  model is to include the sextic term, which is allowed by
requirements of Poincare invariance and existence of the usual Hamiltonian formulation \cite{Adam:2010fg}.
This term is proportional to the square of the topological current, thus it is quadratic
in time derivatives, like the quartic Skyrme term it allows for an usual time dynamics of the system.
It was also noted that inclusion of this term may provide better match with the experimental data of nuclei
\cite{Jackson:1985yz}. Furthermore, the extended Skyrme model with sextic term can be reduced
to the BPS submodel \cite{Adam:2010ds,BPS-other}, which is integrable.

The black hole solutions in the general Skyrme model were studied recently in the asymptotically flat spacetime
\cite{Adam:2016vzf,Gudnason:2016kuu}, the regular gravitating Skyrmions of the BPS submodel were considered in
fixed AdS spacetime  without backreaction \cite{Gudnason:2015dca}.
An interesting observation is that the presence of the quartic Skyrme term is
a necessary condition for the existence of black holes with Skyrmionic hair, there is no hairy black holes
in the BPS submodel in the asymptotically flat spacetime.
However, to our best knowledge, similar study in asymptotically AdS spacetime has not been reported so far.

The main purpose of this work is to extend the consideration of papers \cite{Adam:2016vzf,Gudnason:2016kuu} to the
case of the regular self-gravitating soltions of the general
Skyrme model and the corresponding static hairy black hole solutions in asymptotically AdS$_4$ spacetime.

The paper is organized as follows: in the next section we discuss the generalized
Skyrme model in asymptotically AdS$_4$ space-time, the
spherically symmetric ansatz which we apply to parameterize
the action of the Einstein-Skyrme theory,
the boundary conditions imposed to get regular solution and establish the equations.
The numerical results for the case of regular self-gravitating Skyrmions are presented in
Section 3 where we compare the results to the numerical
solutions of the Skyrme model in asymptotically flat space-time. The properties of the hairy black holes
are discussed in the Section IV. We give our conclusions and remarks in the final section.

%%%%%%%%%%%%%%%%%%%%%%%%%%%%%%%%%%
\section{Generalized Einstein-Skyrme-AdS model}
%%%%%%%%%%%%%%%%%%%%%%%%%%%%%%%%%%

The  general Skyrme model  is a Poincare
invariant, nonlinear sigma model field theory. In a curved spacetime with metric $g_{\mu\nu}$
the Lagrangian of the model
takes the form
\begin{equation}
\label{lag}
\mathcal{L} \equiv a \mathcal{L}_2+ b\mathcal{L}_4+ c \mathcal{L}_6+ \mathcal{L}_0,
\end{equation}
where the quadratic and quartic in derivative terms
\be
\mathcal{L}_2 =  \frac{1}{8}g^{\mu\nu} \mbox{Tr\;} (L_\mu L_\nu), \;\;
\mathcal{L}_4=\frac{1}{64} g^{\mu\nu} g^{\rho\sigma} \mbox{Tr\;} ( [L_\mu, L_\rho] [L_\nu, L_\sigma]) \, ,
\ee
are the sigma model and Skyrme parts, respectively.
The parameters $a,b$ and $c$ are dimensionfull non-negative coupling constants.
The model is expressed in terms if the $su(2)$-valued left chiral current
$L_\mu = U^\dagger \partial_\mu U$, associated with
the $SU(2)$-valued Skyrme field $U = \sigma \cdot {\mathbb I} + i \pi^a \cdot \tau^a$.
The quartet of the fields $(\sigma, \pi^a)$ is restricted to the surface
of the unit sphere, $\sigma^2 + \pi^a \cdot \pi^a =1$. The topological boundary condition on the Skyrme field is that
$U\to \mathbb{I}$ as $r\to \infty$, thus the field is a map $U: S^3 \to S^3$, which belongs
to an equivalence class characterized by the homotopy group $\pi_3(S^3)$. The corresponding topological current is
\begin{equation}
\label{top-curr}
\mathcal{B}^\mu  = -\frac{1}{24 \pi^2 \sqrt{-g}}\varepsilon^{\mu\nu\rho\sigma} \Tr (L_\nu L_\rho L_\sigma)  \, .
\end{equation}
The topological charge corresponds to zeroth component of this current,
thus the winding number is $B=\int d^3 x \sqrt{-g} \mathcal{B}^0$.

Further, the extended Skyrme model \re{lag} includes a sextic term, which is given by the square of the topological
current
\be
\label{sextic}
\mathcal{L}_6= \pi^4 \mathcal{B}_\mu \mathcal{B}^\mu \, .
\ee
It was noted that effectively this term  corresponds
to an inclusion of the repulsive interaction mediated by $\omega$ meson  \cite{Jackson:1985yz} and gives the leading
contribution to the equation of state at high pressure or density \cite{eos}.

The model also includes the potential term $\mathcal{L}_0$, its form can be arbitrary. Here we will use the standard
potential
\be
\label{potential}
\mathcal{L}_0= \frac{{m_\pi}^2}{8}\mbox{Tr\;}\left(U-{\mathbb I}\right),
\ee
where the parameter $m_\pi$ is the pion mass \footnote{Our conventions are slightly different from
usual choice, see e.g. \cite{Shiiki:2005aq}, the matter field sector differs from the standard
one by a factor of one half. Evidently, corresponding rescaling of the cosmological constant $\Lambda$
allows us to recover the latter conventions.}.

Note that in the flat space limit the model \re{lag} can be reduced to the usual Skyrme model in the chiral limit
by setting $c=m_\pi=0$. On the other hand, setting $a=b=0$ reduces the general model to the BPS submodel, with only
the sextic term $\mathcal{L}_6$ and a potential term $\mathcal{L}_0$. The solitons of this model saturate the
topological
bound, they have zero binding energy and, with a usual choice of the potential term \re{potential},
they are compactons \cite{Adam:2010ds}.

Let us consider the extended Einstein-Skyrme model with a cosmological constant
$\Lambda=-3/L^2$. The action is
\be
\label{action}
S=\int d^4 x \sqrt{-g} \left(\frac{R-2\Lambda}{16\pi G}+\mathcal{L}\right).
\ee
where $\mathcal{L}$ is the general Skyrme Lagrangian \re{lag}. The asymptotically flat limit
$\Lambda=0$ of this model was investigated in \cite{Adam:2016vzf,Gudnason:2016kuu}. The other limiting case
$c=0$ corresponds to the standard Skyrme model in the AdS spacetime \cite{Shiiki:2005aq}.

Note that
we can introduce the
dimensionless radial coordinate $\tilde r=\sqrt{\frac{a}{b}}r$
and the effective gravitational coupling constant $\alpha^2=4\pi Ga$. Then only four parameters effectively remain
in the rescaled action:
\be
\label{scale}
a \to 1, \quad b \to 1, \quad c \to \tilde c = \frac{a c}{b^2},
\quad m_\pi^2 \to \tilde m_\pi^2=m_\pi^2 \frac{b}{a^2},
\quad \Lambda \to \tilde \Lambda = \Lambda \frac{b}{a}.
\ee
Alternatively, one can consider the AdS length scale setting $r\to r \sqrt \Lambda$.

The Einstein equations can be obtained from the variation of the action \re{action}
with respect to the metric,
 \begin{eqnarray}
\label{Einstein-eqs}
%R_{\mu\nu}-\frac{1}{2}Rg_{\mu\nu}-\frac{3}{L^2}=8\pi G~ T_{\mu\nu}~,
R_{\mu\nu} -\frac{1}{2}Rg_{\mu\nu} + \Lambda g_{\mu \nu}= 8\pi G T_{\mu\nu}~,
\end{eqnarray}
where the Skyrme stress-energy tensor is defined as
\begin{eqnarray}
\label{Tmunu}
\begin{split}
T_{\mu\nu}&=\frac{2}{\sqrt{-g}}\frac{\delta (\sqrt{-g} {\cal L})}{\delta g^{\mu\nu}}~\\
&=\frac{a}{4}\Tr \left( L_\mu L_\nu -\frac{1}{2}g_{\mu\nu}L_\rho L^\rho \right)+
\frac{b}{16}
\Tr \left(\left[L_\mu,L_\rho \right]\left[L_\nu,L^\rho \right]-\frac{1}{4}g^{\mu\nu}
\left[L_\rho,L_\sigma \right]\left[L^\rho,L^\sigma \right]\right)\\
&+\frac{c}{64}\left(\Tr\left(L_\mu\left[L_\alpha,L_\sigma\right]\right)\Tr\left(L_\nu\left[L^\alpha,L^\sigma\right]\right)
-\frac{1}{6}g_{\mu\nu}\Tr\left(L_\alpha\left[L_\beta,L_\sigma\right]\right)
\Tr\left(L^\alpha\left[L^\beta,L^\sigma\right]\right)
\right)
\end{split}
\end{eqnarray}
Apart from (\ref{Einstein-eqs}), one should also consider the matter field equations, which
are found from the variation of the action with respect to the Skyrme field $U$.

Hereafter we restrict our consideration to the static spherically symmetric configurations in the sector $B=1$.
Then the Skyrme field can be parametrized by the rotationally invariant hedgehog ansatz
\be
\label{hedgehog}
U({\bf r})= e^{if(r) \hat r^a\cdot \tau^a } = \cos f(r) + i\sin f(r) \hat r^a\cdot \tau^a
\ee
where the Skyrmion profile function $f(r)$ is a real function of the radial variable
with the topological boundary conditions $f(0)=\pi$,
$f(\infty)=0$, and $\hat r^a$ the usual unit vector in $\mathbb{R}^3$. We also make use of the
static, spherically symmetric asymptotically AdS  metric
\be
\label{metric}
ds^2=\sigma^2(r)N(r)dr^2-\frac{dr^2}{N(r)}-r^2\left(d\theta^2+\sin^2\theta d\phi^2\right)
\ee
with the  metric function $N(r),\sigma(r)$ depending only on $r$. It is convenient to define
\be
\label{ADM}
N(r)=1-\frac{2M(r)}{r}-\frac{\Lambda r^2}{3} \, ,
\ee
then in the limit $r\to\infty$ the metric function $M(r)$ approaches the
so called Arnowitt-Deser-Misner (ADM) mass of the solution.

Substituting \re{hedgehog},\re{metric} into \re{action} and varying it with respect to
$N(r)$, $\sigma(r)$ and $f(r)$ we obtain the following system of coupled  equations:
\be
\label{eqs}
\begin{split}
N_{r}=&-\frac{\alpha ^2}{2} \biggl[a \left(N r f_r^2+\frac{2 \sin ^2{f}}{r}\right)+b \left(\frac{2 N f_r^2 \sin ^2{f}}{r}+\frac{\sin ^4{f}}{r^3}\right)+\frac{c N f_r^2 \sin ^4{f}}{r^3}+\frac{1}{2} m_{\pi }^2 r \sin ^2\frac{f}{2}\biggr]\\
&+\frac{1-N}{r}-\Lambda  r,\\
\sigma_r=&\frac{1}{2} \alpha^2 \sigma  f_r^2\left(a r+\frac{ 2b \sin^2 f}{r}+\frac{c \sin^4f }{r^3}\right),\\
f_{rr}=&\frac{1}{a+\frac{2 b \sin ^2{f}}{r^2}+\frac{c \sin ^4{f}}{r^4}}\biggl\{a \biggl[\frac{\sin {2 f}}{N r^2}-f_r \left(\frac{N_r}{N}+\frac{\sigma _r}{\sigma }+\frac{2}{r}\right)\biggr]\\
&+b \left(\frac{2 \sin ^3{f} \cos {f}}{N r^4}+f_r \left(-\frac{2 \sin ^2{f}}{r^2}\frac{N_r}{N}-\frac{2 \sin ^2{f}}{r^2}\frac{\sigma_r}{\sigma}\right)-\frac{f_r^2 \sin {2 f}}{r^2}\right)\\
&+c \left(f_r \left(-\frac{\sin ^4{f}}{r^4}\frac{N_r}{N}+\frac{2 \sin ^4{f}}{r^5}-\frac{\sin ^4{f}}{r^4}\frac{\sigma_r}{\sigma}\right)-\frac{2 f_r^2 \sin ^3{f} \cos {f}}{r^4}\right)+\frac{m_{\pi }^2 \sin {f}}{2 N}\biggr\}.
\end{split}
\ee

Similar to the case of the general Skyrme model in asymptotically flat space \cite{Adam:2016vzf},
the system of equations \re{eqs} is invariant with respect to the scaling transformation $\sigma \to \lambda \sigma$.
Thus, the function $\sigma$ can be rescaled to set $\sigma \to 1$ as $r\to \infty$, furthermore this function
can be eliminated from the equations completely since the third  of these
equations depends only on ratio $\frac{\sigma_r}{\sigma}$, given by the second equation. Thus,
\be
\label{sigmint}
\sigma(r)=\sigma(0) \exp\left[\frac{\alpha^2}{2} \int_0^r{f_r^2\left(a r+\frac{ 2b \sin^2 f}{r}+
\frac{c \sin^4f }{r^3}\right)dr}\right] \, ,
\ee
where $\sigma(0)$ is an integration constant. However, it is convenient to keep the metric function $\sigma$
in the numerical calculations.

The equations \re{eqs} can be solved numerically when we impose the
appropriate boundary conditions.
In the case of regular self-gravitating solitons, the boundary conditions at the origin and on the spacial
asymptotic follows from the conditions of finiteness of the energy functional and topology of the
Skyrme field in the sector of degree one:
\be
\label{bcond}
\begin{split}
f(0)=\pi,\quad N(0)=1, \quad f(\infty)=0, \quad \sigma(\infty)=1.
\end{split}
\ee

%%%%%%%%%%%%%%%%%%%%%%%%%%%%%%%%%%
\section{Gravitating solitons in the generalized Einstein-Skyrme-AdS model}
%%%%%%%%%%%%%%%%%%%%%%%%%%%%%%%%%%

Solutions of the system \re{eqs} are constructed numerically using the 8th order shooting algorithm based on the
Dormand-Prince modification of the Runge-Kutta method \cite{RKDP}. They are also verified by
boundary-value problem solving routines of open source package bvpSolve \cite{BVP}, the relative
error is lower than $10^{-9}$.

\begin{figure}[hbt]
\lbfig{f1}
\begin{center}
\includegraphics[height=.26\textheight, trim = 60 20 80 50, clip = true]{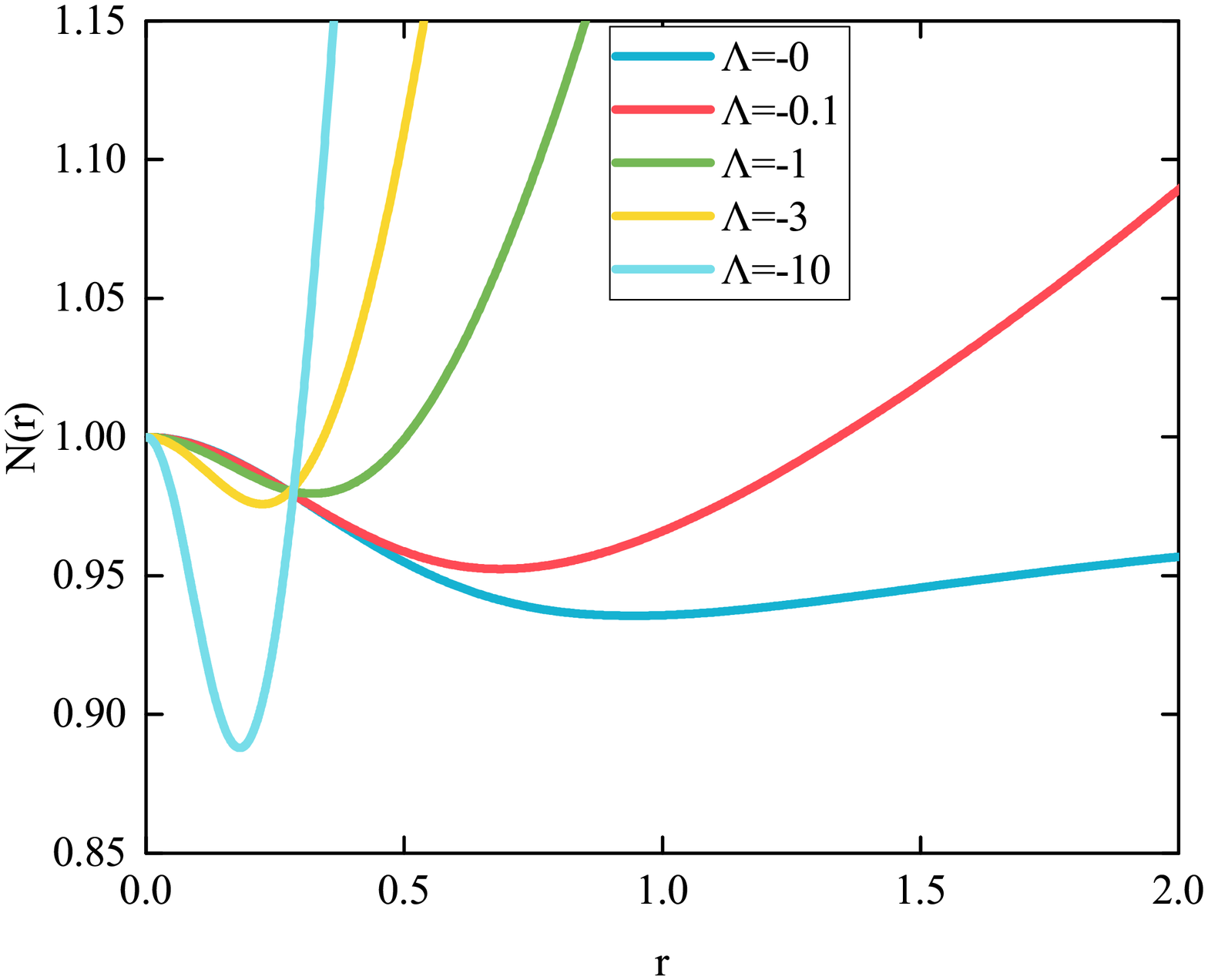}
\includegraphics[height=.26\textheight, trim = 60 20 80 50, clip = true]{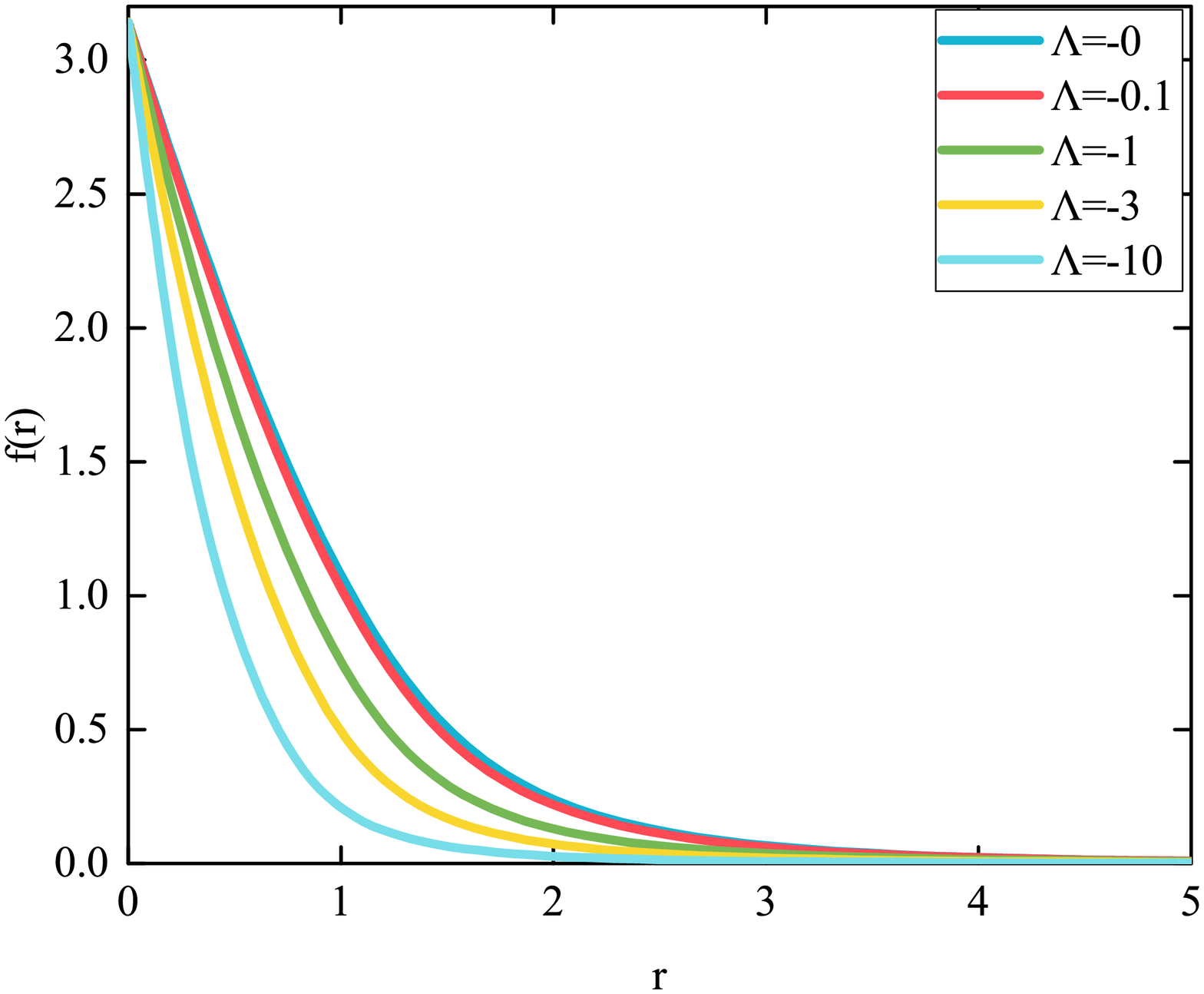}
\includegraphics[height=.26\textheight, trim = 60 20 80 50, clip = true]{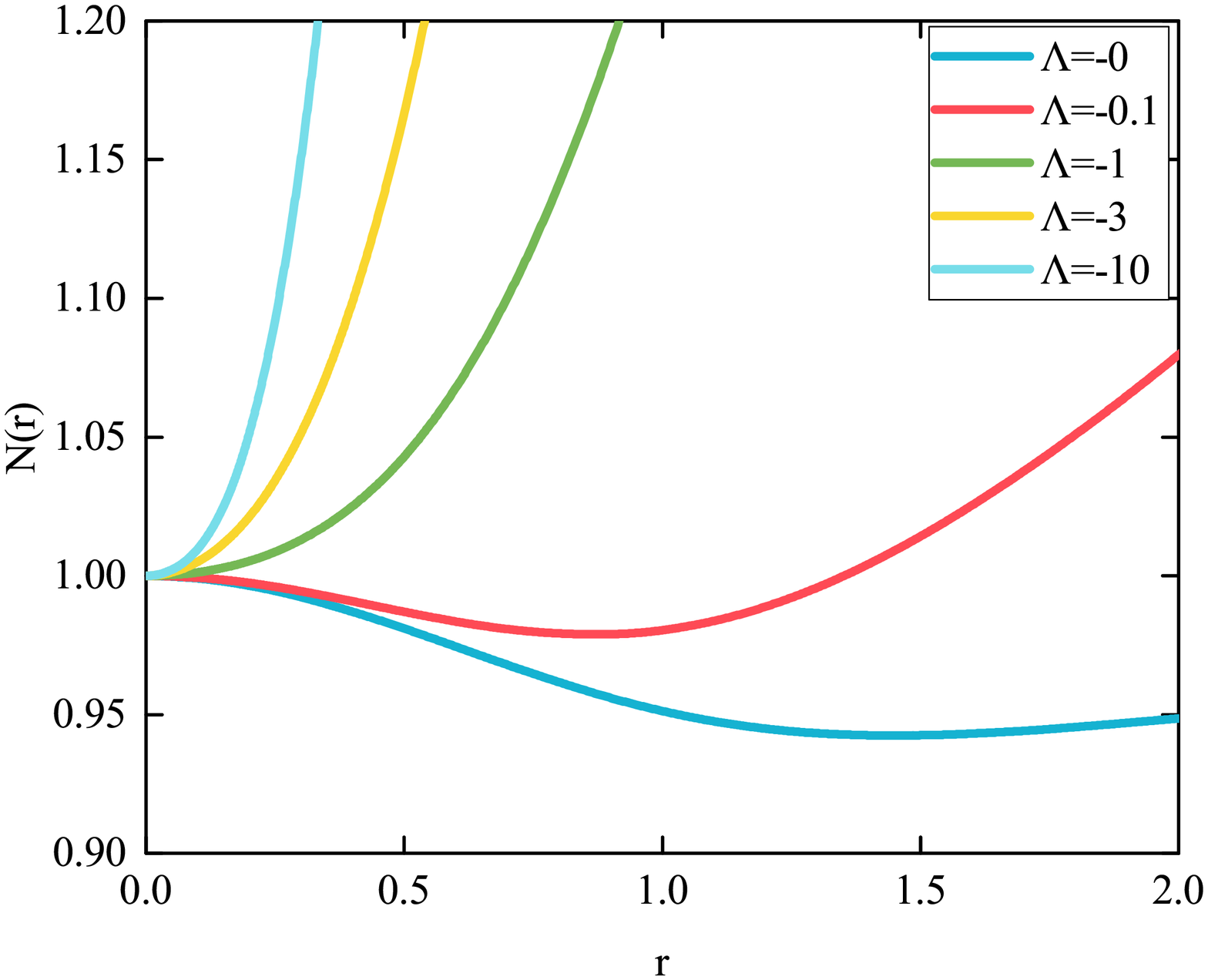}
\includegraphics[height=.26\textheight, trim = 60 20 80 50, clip = true]{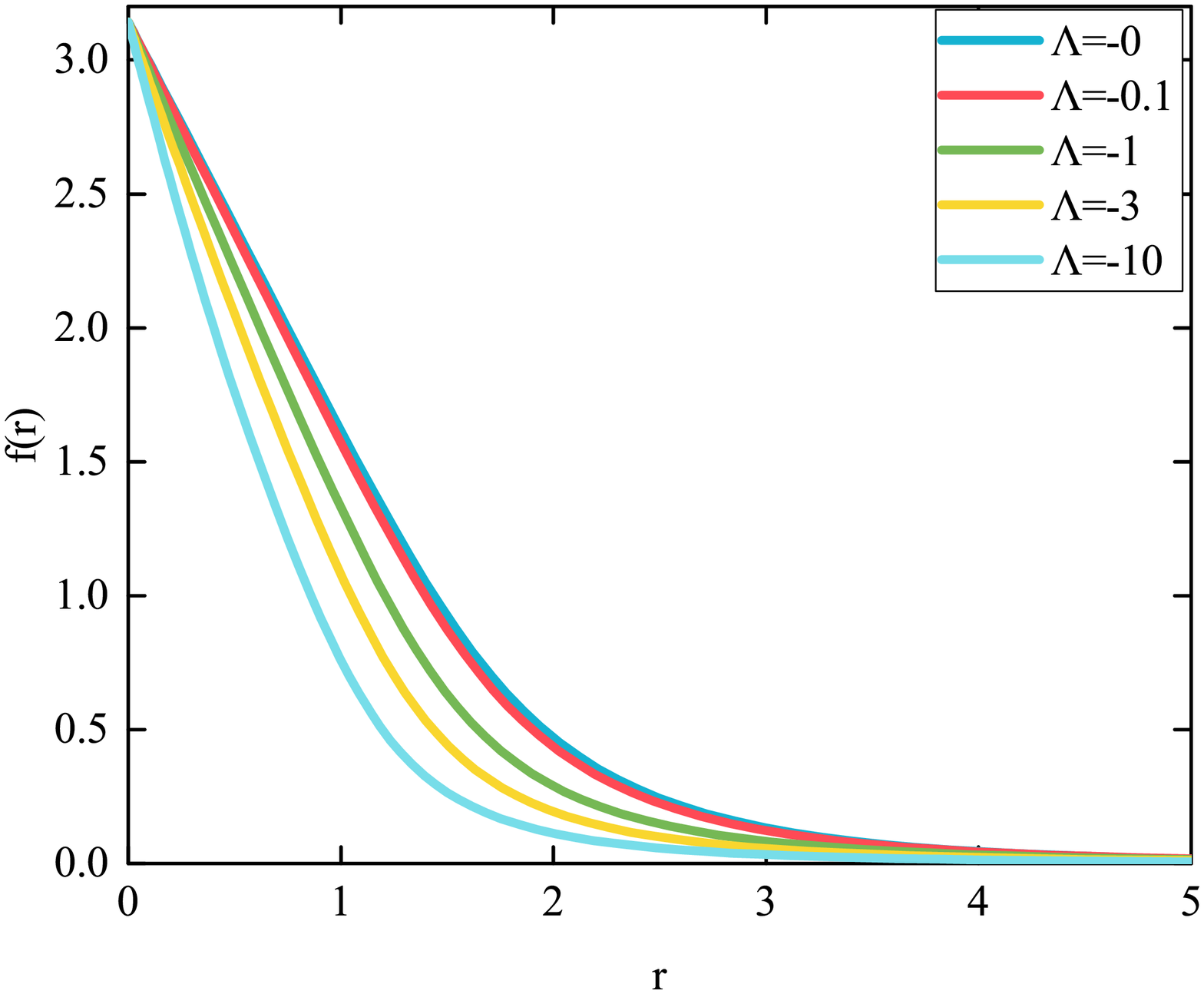}
\end{center}
\caption{\small
The metric function $N(r)$ (left column) and the Skyrmion  profile function $f(r)$ (right column)
of the self-gravitating Skyrmion are plotted as functions of the  radial coordinate $r$
for $c=0$ (upper rows) and $c=1$ (lower row), for $a=b=1$, $\alpha=0.12$, lower branch.
}
\end{figure}

First, we consider the regular solutions of the system \re{eqs}. The system of equations
\re{eqs} depends on six parameters: four coupling constants in the matter fields sector,
$a, ~b, ~c, ~m_\pi$, the effective gravitational coupling constant $\alpha$ and the
cosmological constant $\Lambda$. Evidently, the corresponding pattern of solutions is extremely rich.
It can be simplified if we note that the rescaling of the parameters of the model \re{scale}
allows us to fix $a=b=1$. Further, without loss of generality,
we can take $m_\pi=1$, so for the regular solutions which approach asymptotically an
 AdS$_4$ background, only three parameters $c, ~\alpha$ and $\Lambda$ remain.

\begin{figure}[hbt]
\lbfig{f2}
\begin{center}
\includegraphics[height=.26\textheight, trim = 60 20 80 50, clip = true]{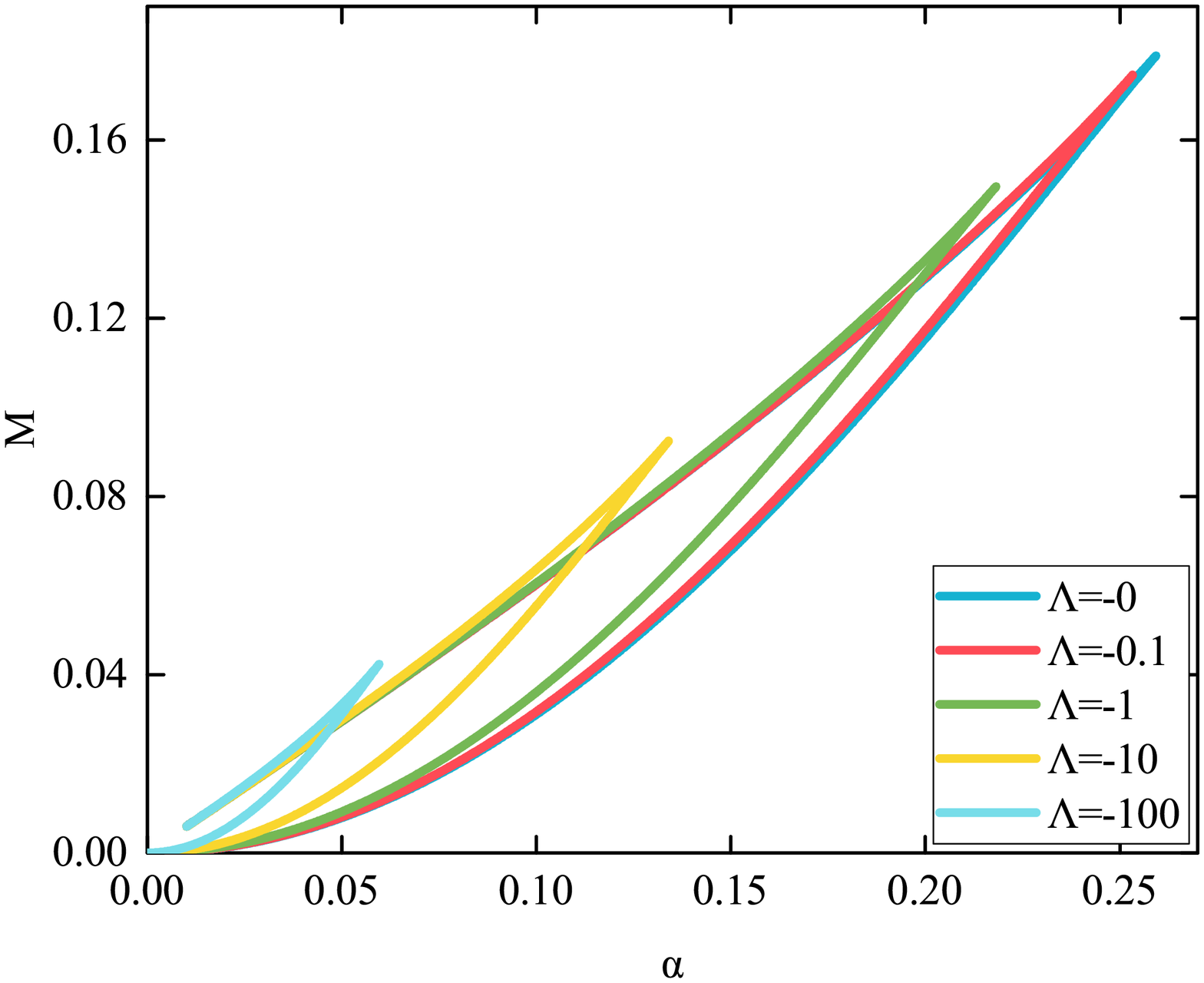}
\includegraphics[height=.26\textheight, trim = 60 20 80 50, clip = true]{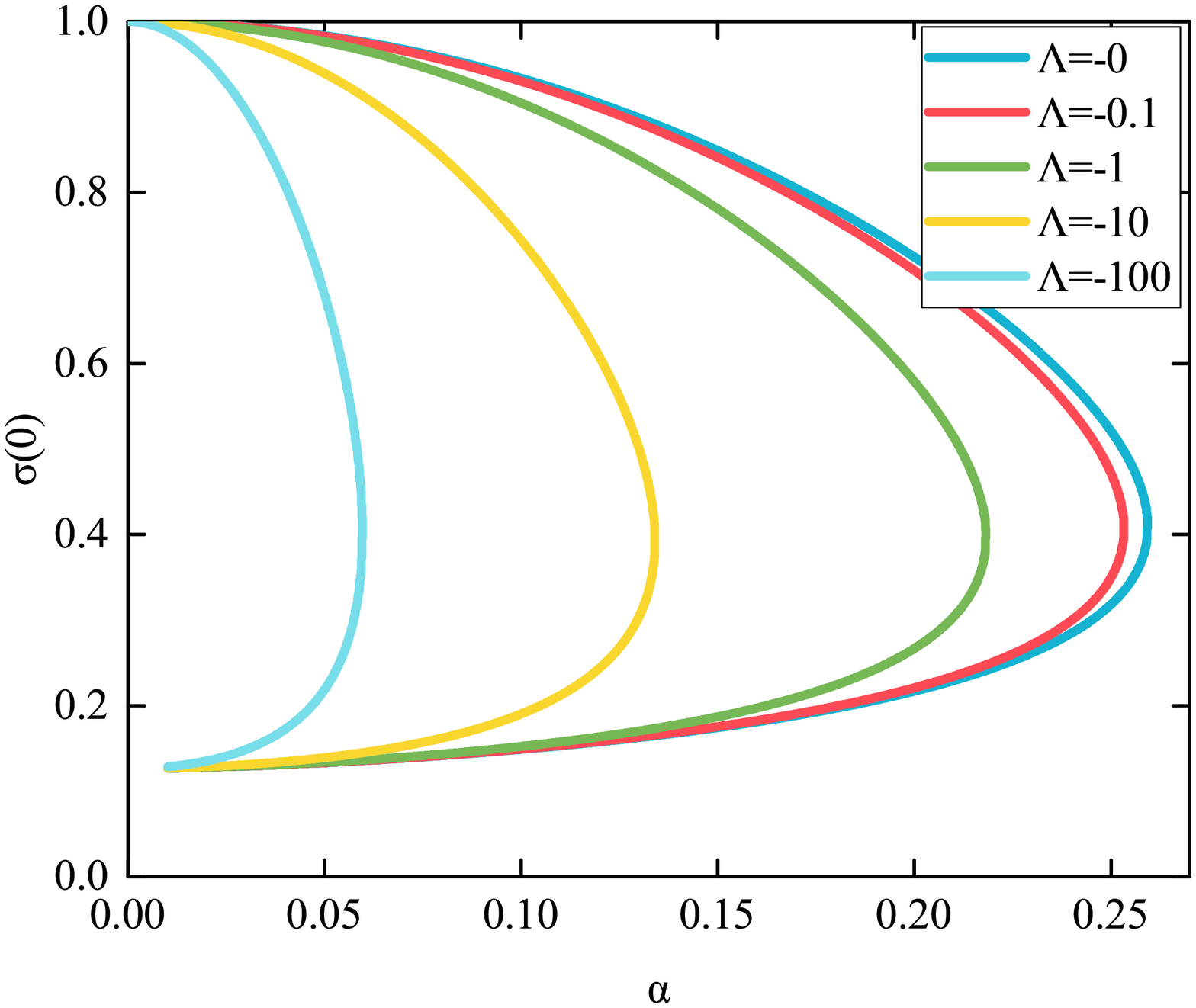}
\includegraphics[height=.26\textheight, trim = 60 20 80 50, clip = true]{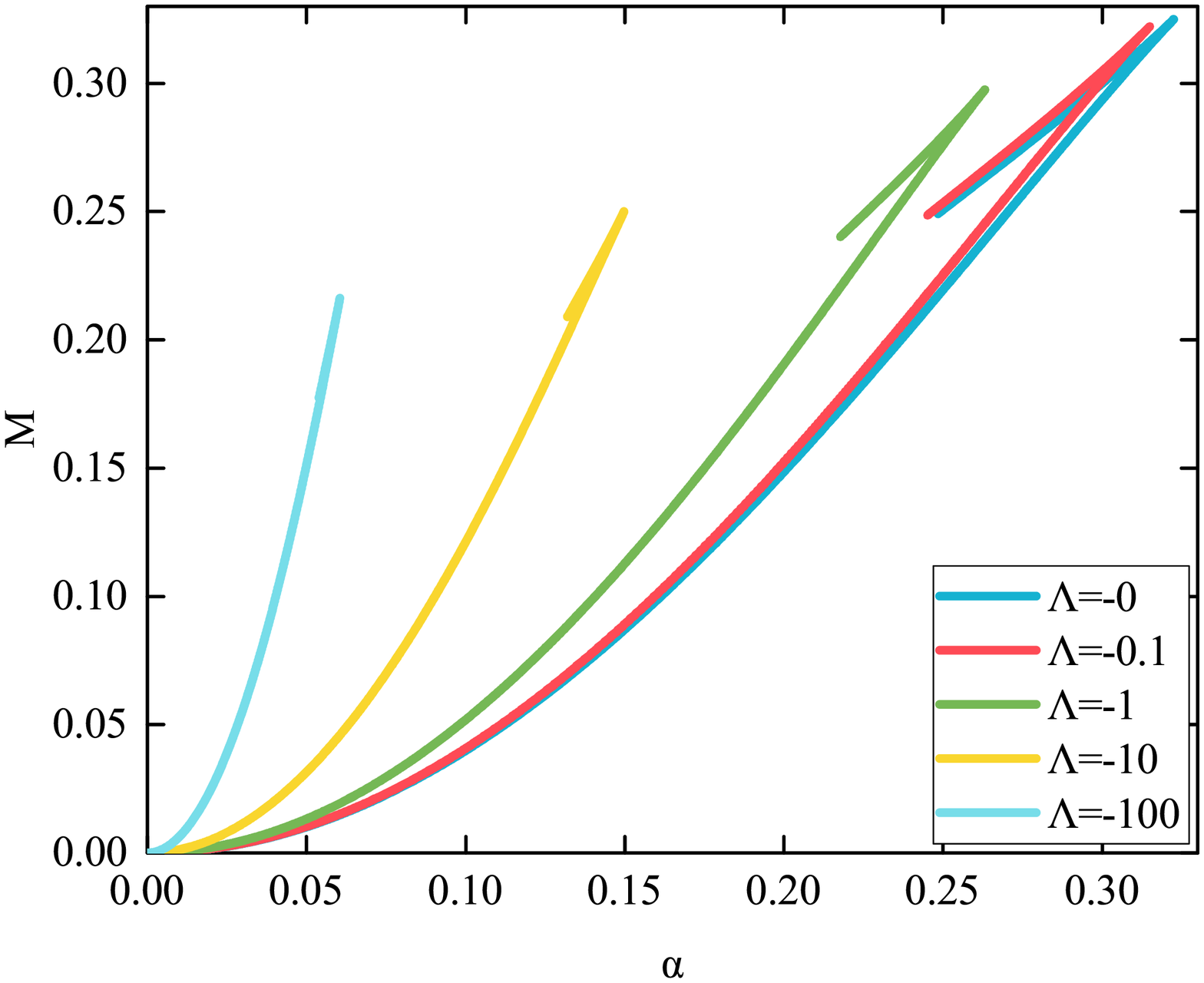}
\includegraphics[height=.26\textheight, trim = 60 20 80 50, clip = true]{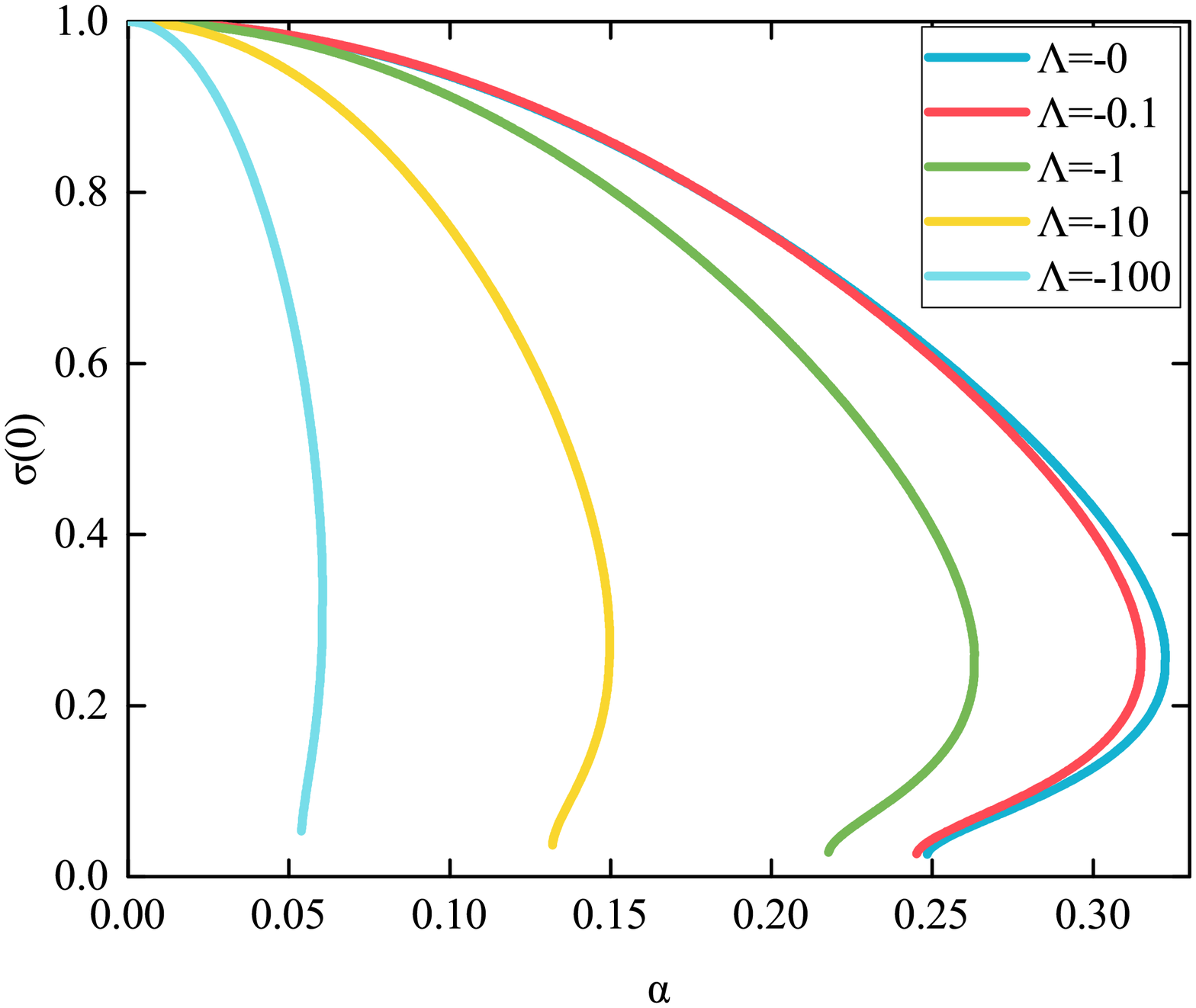}
\end{center}
\caption{\small
The value of  the metric function $\sigma$  of the self-gravitating Skyrmions
at the origin, $\sigma(0)$,  (left) and the ADM mass of the corresponding
solutions (right) are plotted as functions of the coupling constant $\alpha$
for $a=b=m_\pi=1$ and a few values of $\Lambda$.  The upper and lower rows correspond to the
cases $c=0$ and $c=1$, respectively.
}
\end{figure}

Properties of the self-gravitating solutions of the general model \re{action}
in the asymptotically flat case has been studied before  \cite{Adam:2016vzf},
thus we will extend our consideration to the case of $|\Lambda| \neq 0$.

First, we note that,
similar to the case of the usual Einstein-Skyrme theory with a negative cosmological constant \cite{Shiiki:2005aq},
the branch of self-gravitating solutions
emerges smoothly from  the corresponding configuration in the fixed AdS background for any values of
$c$ and $\Lambda$ as $\alpha$ start to increase.

On this branch, in the limiting case of the
usual Skyrme model, $c=0$, for a given $\alpha$ the metric function $N(r)$ develops a minimum at some $r_0 \neq 0$.
This minimum moves towards the origin as $|\Lambda|$ increases, see Fig.~\ref{f1}. The
depth of the minimum non-monotonically depends on $|\Lambda|$, it becomes least pronounced for $\Lambda= -1$.

In a contrast, in the general model
at $c=1$, the minimum of the metric function $N(r)$ at $r_0 \neq 0$ for a given $\alpha$ exist
for relatively small values of $|\Lambda|<1$.  As $|\Lambda|$
becomes bigger, there is a single minimum of this function at the origin. As expected, the
soliton core  shrinks with increase of $|\Lambda|$ for any values of $c$.

Investigating the dependence of the solutions
on the effective gravitational coupling constant $\alpha$, we see that
both the value of the metric function $\sigma(0)$ at the origin and the minimum of the metric function
$N(r)$, if it exists, monotonically decrease as $\alpha$ increases, as seen in Fig.~\ref{f2}.
At the same time, the ADM mass of the solutions monotonically increases.
However the rate of decrease of $\sigma(0)$, as well as the rate of increase of the ADM mass, incrementally
grows as $|\Lambda|$ is getting large, both for $c=0$ submodel and for general model with $c=1$.

\begin{figure}[hbt]
\lbfig{f3}
\begin{center}
\includegraphics[height=.28\textheight, trim = 60 20 80 50, clip = true]{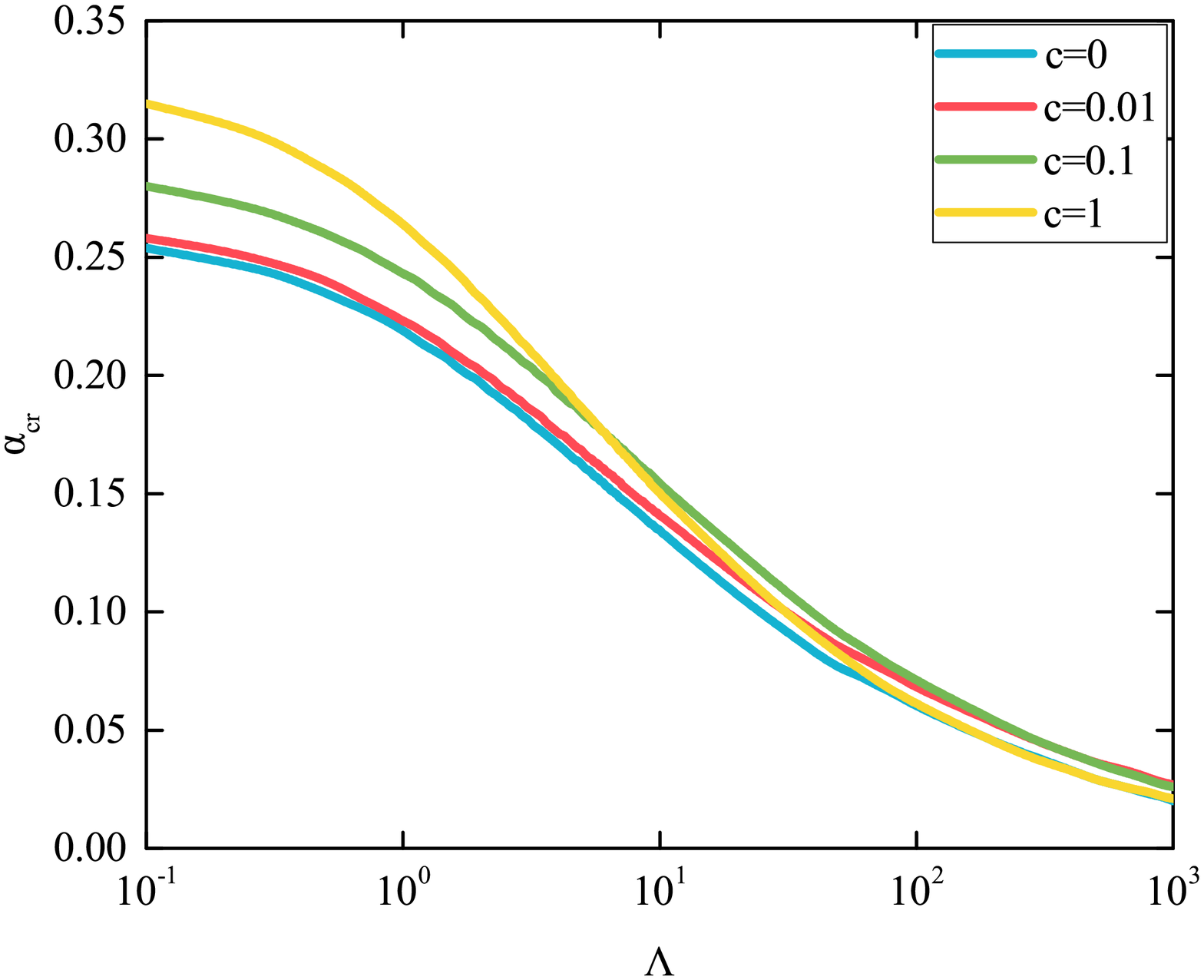}
\end{center}
\caption{\small
The critical value of the gravitational coupling $\alpha_{cr}$ is shown as function of $|\Lambda|$
for some set of values of $c$.
}
\end{figure}

This branch extends up to some maximal value of effective gravitational coupling
$\alpha_{cr}$, beyond which the gravitational interaction becomes too strong for gravitating Skyrmions
to exist. The value of $\alpha_{cr}$ increases with increasing $c$ and it decreases with increasing
$\Lambda$, see Fig.~\ref{f3}, our numerical results clearly indicate that the regular solutions of the model
\re{action} exist for any arbitrary large values of $|\Lambda|$.

At $\alpha=\alpha_{cr}$ this branch bifurcates with the
second, higher-mass branch of regular solutions of the model \re{action}, which extends backwards,
see Fig.~\ref{f2}. Our results confirm the observation that the gravitating solitons of the $c=0$
usual Einstein-Skyrme exist for all values of the coupling constant $\alpha \to 0$
\cite{Adam:2016vzf,Gudnason:2016kuu}. In this limit for any values of $\Lambda$
the rescaled configuration  approaches the lowest
Bartnik-McKinnon  solution of the $SU(2)$ Einstein-Yang-Mills theory \cite{Bartnik:1988am}.

The pattern of evolution of the upper branch regular solutions for
non-zero values of the coupling constant $c$ is different since the scaling properties of the
general model \re{action} then are different. We confirm the observation \cite{Adam:2016vzf}
that for any $c \neq 0$ this branch exist only for $\alpha>\alpha_{min}$, at which $\sigma(0) \to 0$.
The value of $\alpha_{min}$ decreases with increasing $\Lambda$, cf. Fig.~\ref{f2}. Along this branch the ADM mass
of the solutions monotonically decreases, the branch us getting shorter as $\Lambda$ increases.

\begin{figure}[hbt]
\lbfig{f4}
\begin{center}
\includegraphics[height=.26\textheight, trim = 60 20 80 50, clip = true]{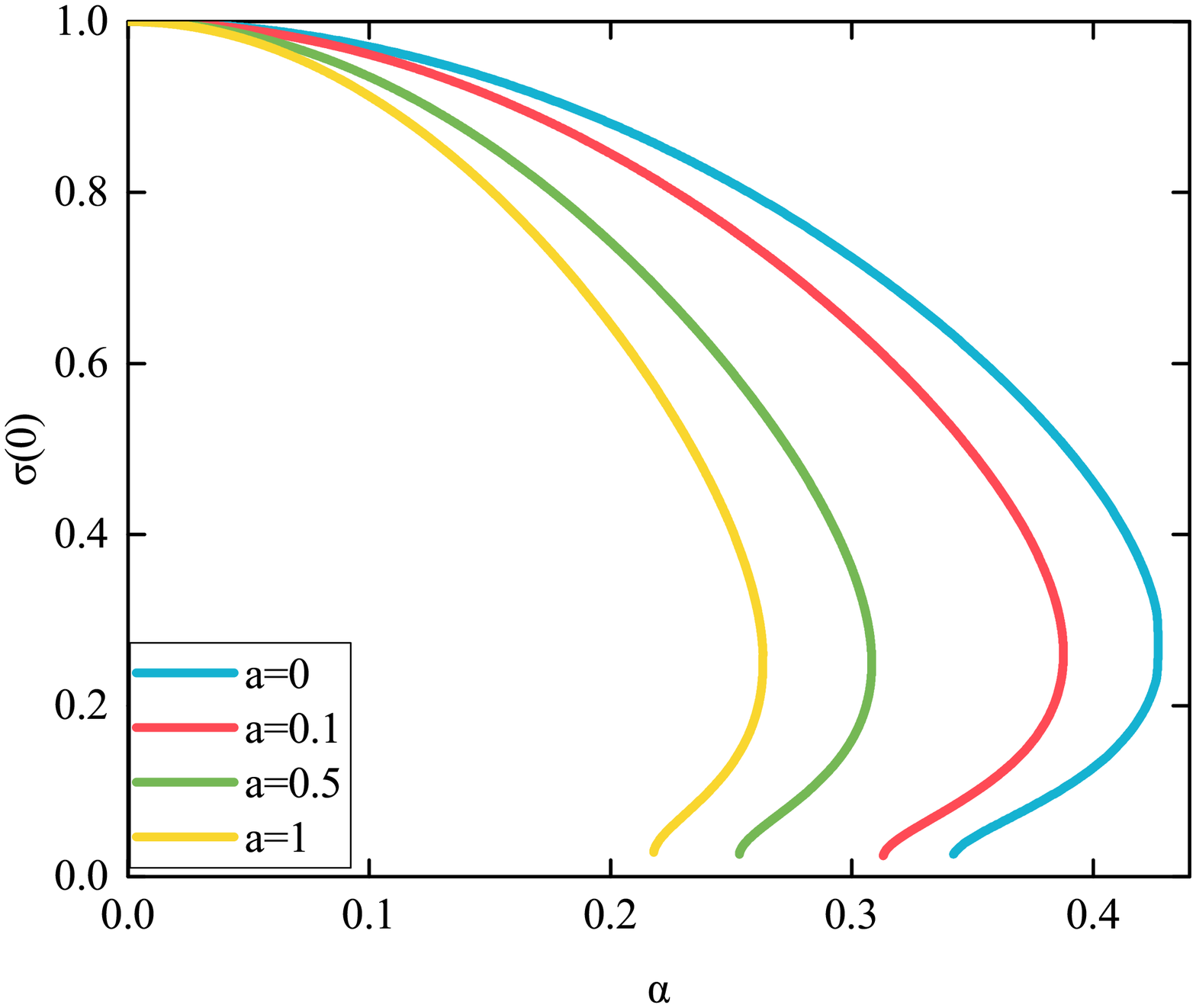}
\includegraphics[height=.26\textheight, trim = 60 20 80 50, clip = true]{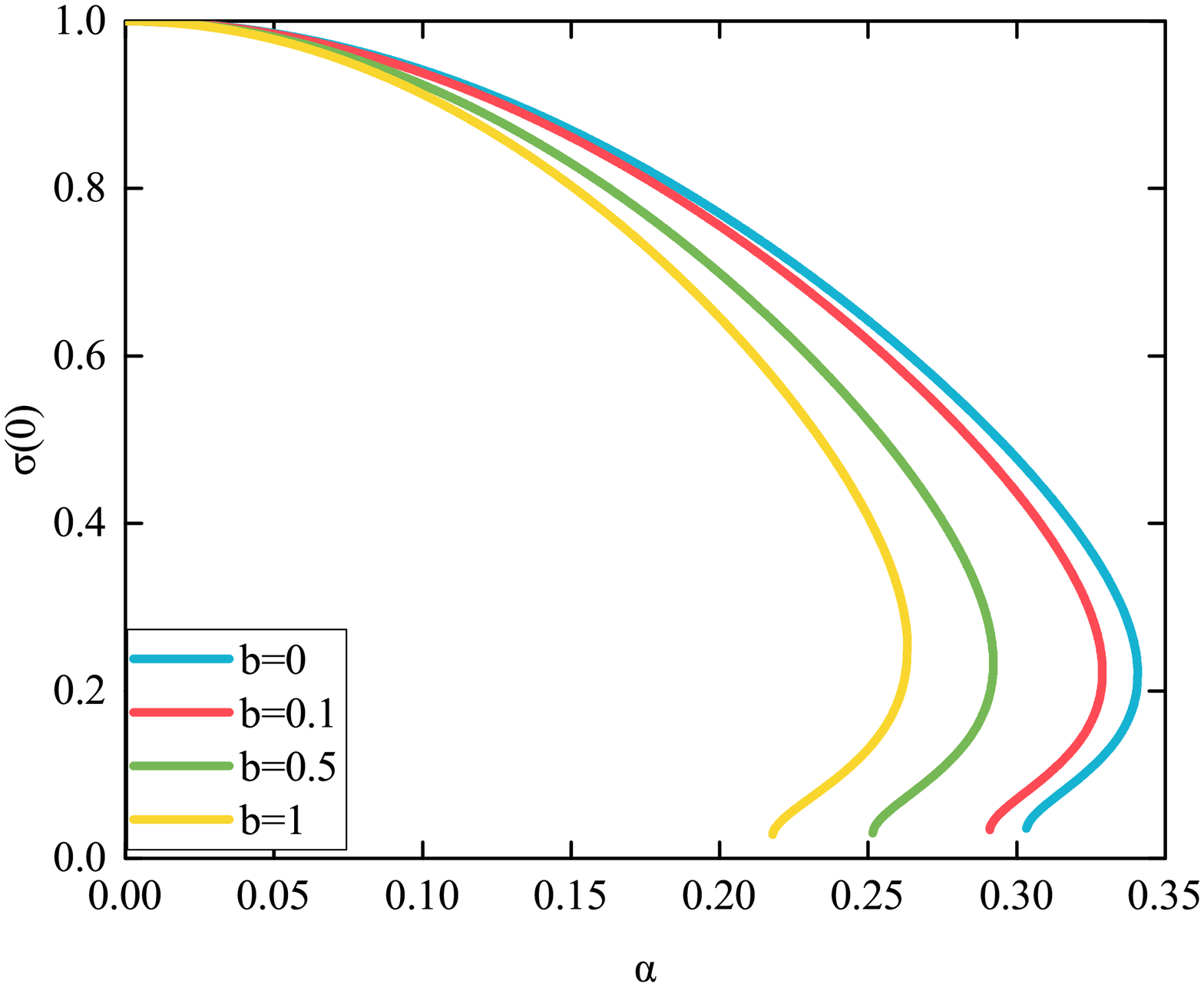}
\end{center}
\caption{\small
The value of the metric function $\sigma(0)$ of the self-gravitating Skyrmions
at the origin is shown as function of the  gravitational coupling constants $\alpha$
for some set of values of the coupling constant $a$ at $b=1$ (left plot) and for some set of values
of the coupling constant $b$ at $a=1$ (right plot),
at $c=1$ and $\Lambda=1$.
}
\end{figure}

We can also consider limiting submodels $\mathcal{L}_0+\mathcal{L}_4+\mathcal{L}_6$,
$\mathcal{L}_0+\mathcal{L}_2+\mathcal{L}_6$ and $\mathcal{L}_0+\mathcal{L}_6$, for which scaling properties
are different, setting the values of the parameters $a=0,~ b=0$ and $a=b=0$ respectively.
The case $b=0$, however, reveals no interesting features since the behavior of the solutions
actually follows the pattern we observed in general case. As in the asymptotically flat case,
the solutions of the $a=0$ submodels are solitons with compact support for any values of $\Lambda$, the
size of the compacton is decreasing as $|\Lambda|$ grows. Qualitatively, the pattern of evolutions of these
regular self-gravitating Skyrmions as other parameters of the $a=0$ submodel are varying,
is similar to the general case of the full $a \neq 0$ model, in particular, as both parameters $a$ and $b$ are set to
zero, the second, backward branch ceases to exist.

Note that for a fixed value of $c$, both the value of $\alpha_{cr}$ and the ADM mass of the
configuration increase as  $a$ decreases. Also the maximal
value of the effective gravitational coupling increases as  $b$ decreases, see Fig.~\ref{f4}.

%%%%%%%%%%%%%%%%%%%%%%%%%%%%%%%%%%
\section{Hairy black holes in the generalized Einstein-Skyrme-AdS model}
%%%%%%%%%%%%%%%%%%%%%%%%%%%%%%%%%%
It was pointed out recently that the $b=0$ submodels without the Skyrme term in the asymptotically flat 3+1 dim
spacetime do not possess hairy black hole solutions, even
though they support both the static flat space solutions and the regular self-gravitating Skyrmions
\cite{Adam:2016vzf,Gudnason:2016kuu}. In the general model, similar to the usual
Einstein-Skyrme $\mathcal{L}_2+\mathcal{L}_4$ submodel, there are branches of two types, the $\alpha$-branches which
exist at fixed value of the event horizon radius $r_h$, and $r_h$-branches for a fixed value of the gravitational
coupling $\alpha$  \cite{Droz:1991cx,Bizon:1992gb}. The black holes with Skyrmionic hair cannot be
arbitrarily large, there is a secondary $r_h$-branch, extending backward towards $r_h=0$ for  sufficiently small, or zero values of $c$
\cite{Adam:2016vzf,Gudnason:2016kuu}. In other words, the $r_h$-branch yields a link between the regular solutions on the lower and upper
branches at given $\alpha$.

Let us now consider the black hole solutions with general Skyrmionic hair in asymptotically AdS$_4$ spacetime.
The area of the spherically symmetric horizon is $A=4\pi r_h^2$ with the entropy $S=A/4$
and the surface gravity of static black hole can be evaluated from the metric functions on the horizon:
\be
\kappa^2= -\frac{1}{4} g^{tt}g^{rr}(\partial_r g_{tt})^2
\ee
The Hawking temperature of the static Schwarzschild-like black hole is related to the surface gravity as $T_H=\kappa/(2\pi)$, in
the model with cosmological constant $\Lambda$ without the matter field it is
\be
\label{TAdS}
T_H=\frac{1}{4\pi r_h}\left(1- \Lambda r_h^2\right)
\ee

Considering the black holes with general Skyrme hair, we will imply the spherically metric in the usual
AdS-Schwarzschild coordinates \re{metric}
with the metric function $N(r)$ related to the mass function $M(r)$ via \re{ADM}.
Then the Hawking temperature of the hairy black
hole can be evaluated as
\be
\label{hawking}
T=\frac{1}{4\pi}\sigma\left(r_h\right)N'\left(r_h\right)
\ee

As in the case $\Lambda=0$, the boundary conditions on the
matter field and on the metric function must secure the zeroth law of the black holes:
the invariants like the surface gravity, and the Kretschmann scalar
$K=R_{\mu\nu\rho\sigma}R^{\mu\nu\rho\sigma}$  must be finite on the event horizon. In the case under consideration
the asymptotic expansion of the fields in powers of $r-r_h$ in the near-horizon area yields

\begin{figure}[hbt]
\lbfig{fMlambda}
\begin{center}
\includegraphics[height=.26\textheight, trim = 60 20 80 50, clip = true]{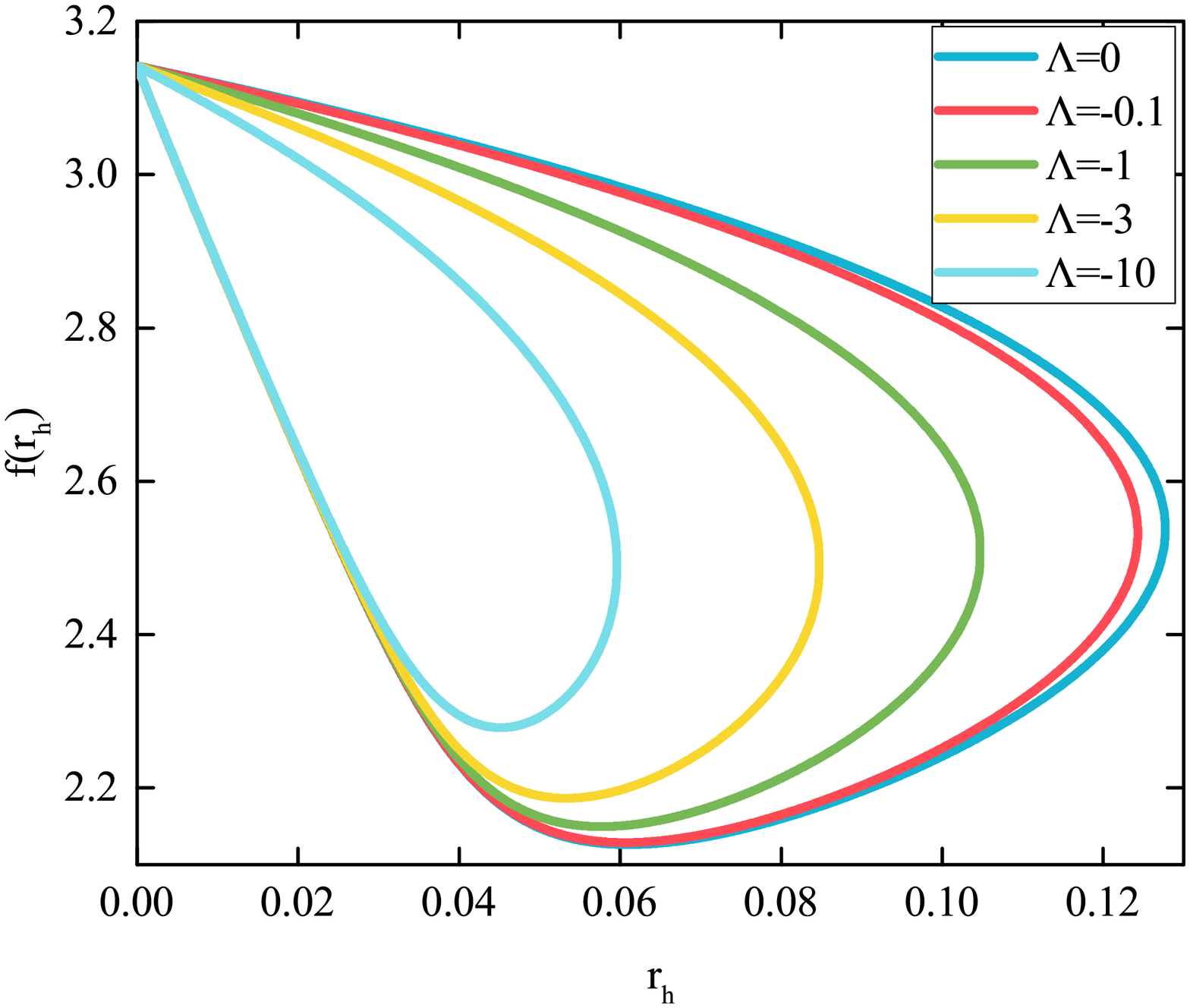}
\includegraphics[height=.26\textheight, trim = 60 20 80 50, clip = true]{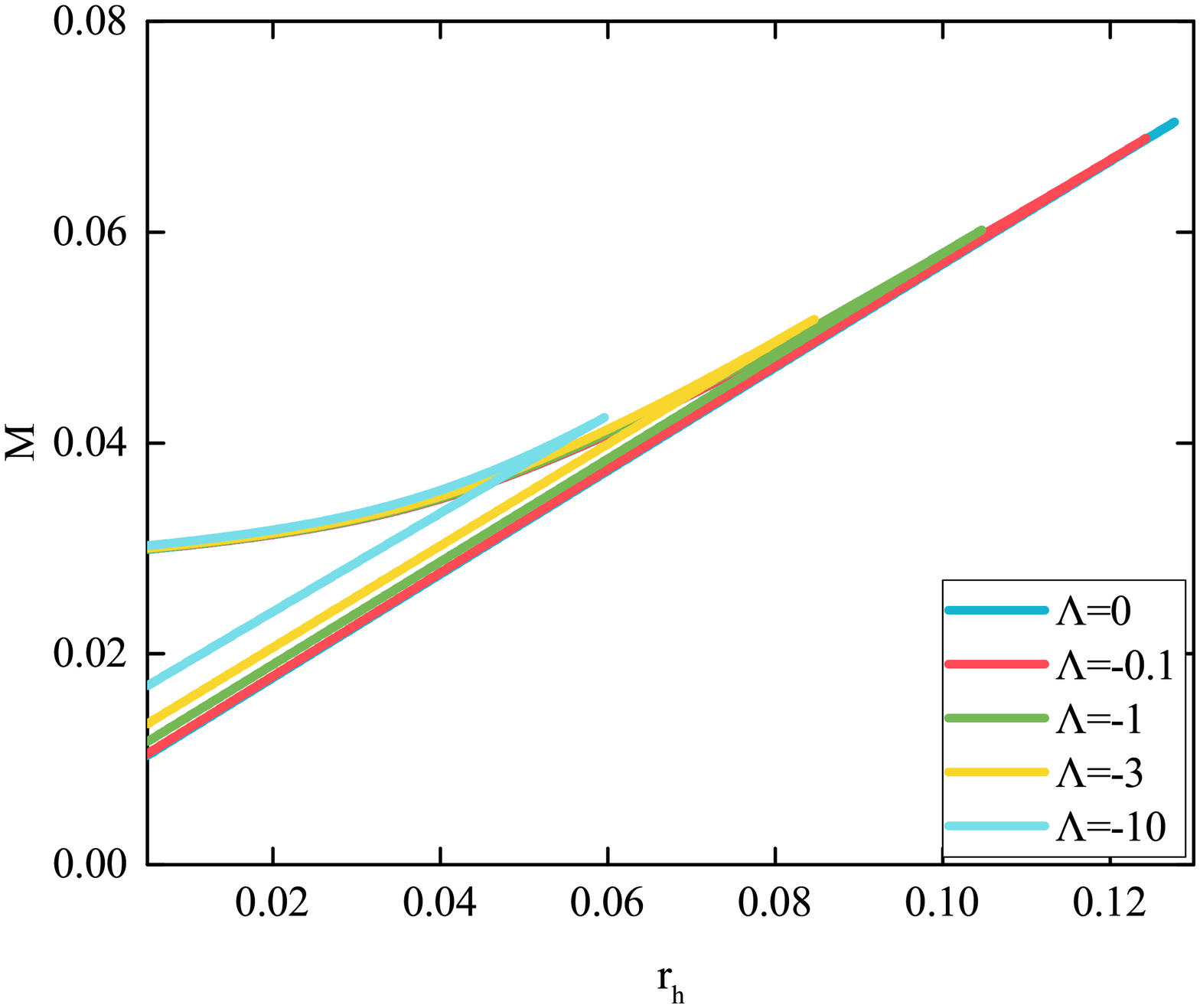}
\includegraphics[height=.26\textheight, trim = 60 20 80 50, clip = true]{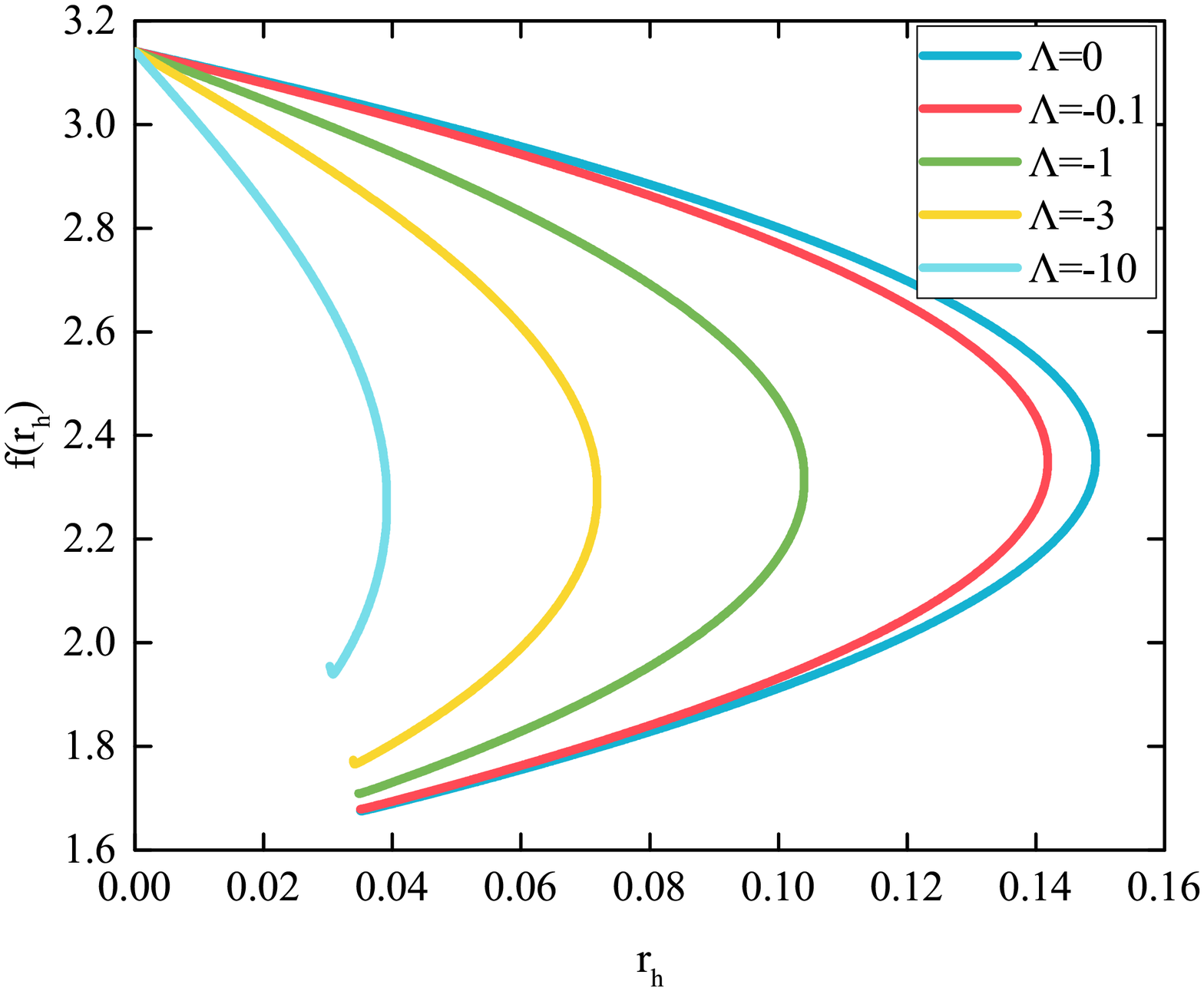}
\includegraphics[height=.26\textheight, trim = 60 20 80 50, clip = true]{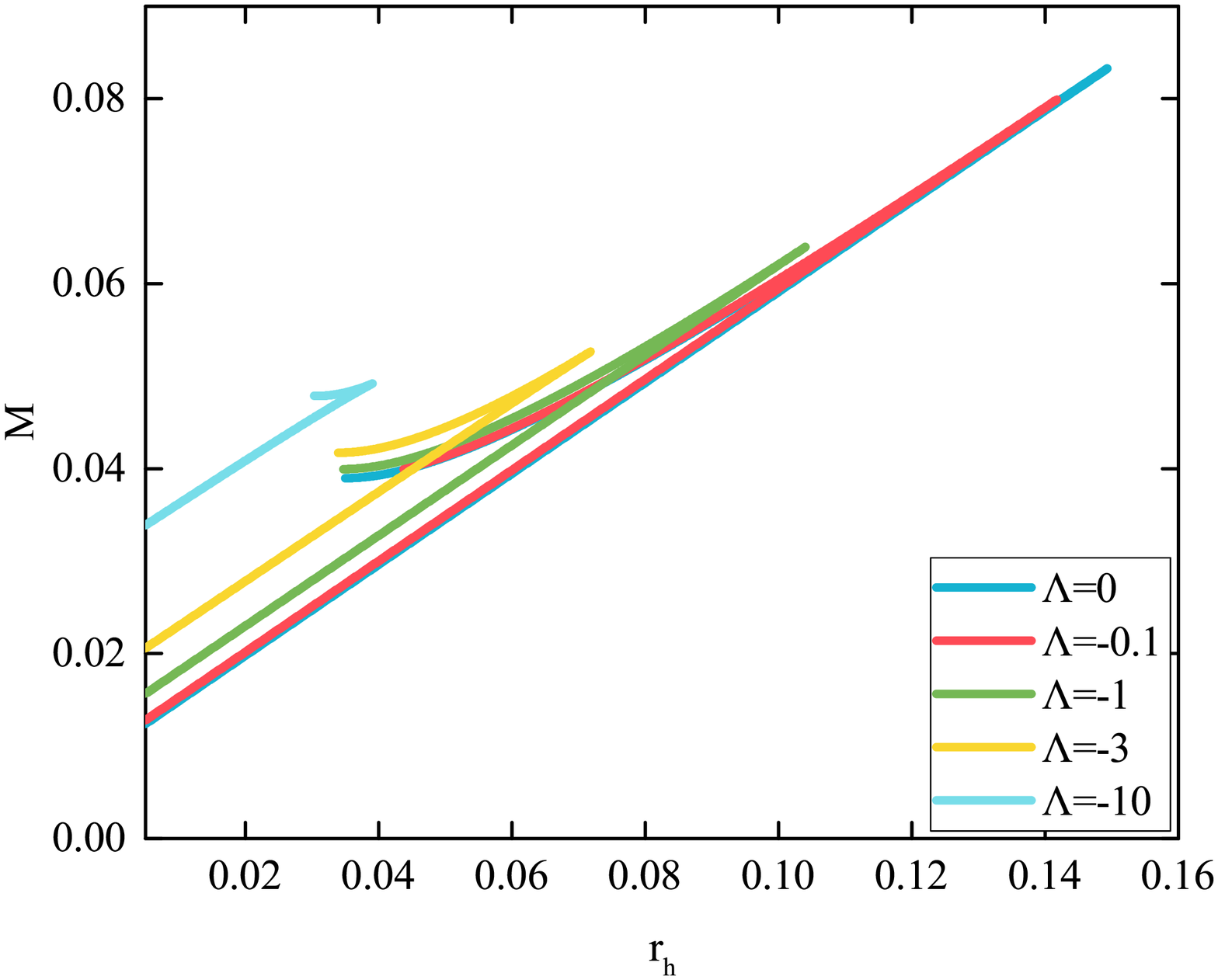}
\end{center}
\caption{\small
The value of the Skyrme field profile function $f(r_h)$ at the event horizon (left column)
and  the ADM mass of the solution (right column) are plotted as functions of the horizon radius $r_h$ for
$\alpha=0.05$, $a=b=m_\pi = 1$, $c=0$ and $c=1$ (upper and lower rows, respectively),
and some set of values of the cosmological constant $\Lambda$.
}
\end{figure}

\be
\label{horexp1}
\begin{split}
N= &N_1 \left(r-r_h\right)+O\left(r-r_h\right)^2,\\
\sigma=&\sigma_h+\frac{\alpha^2 \sigma_h J^2}{8r_h^3HN_1^2}(r-r_h)+O\left(r-r_h\right)^2.\\
f= &f_h+\frac{J}{2HN_1}(r-r_h)+O\left(r-r_h\right)^2,\\
&
\end{split}
\ee
where where we introduced the shorthand notations
\be
\label{horexp2}
\begin{split}
N_1=&\frac{1}{r_h}-\Lambda  r_h-\frac{\alpha ^2}{2} \left(m_{\pi }^2 r_h
\left(1-\cos f_h\right)+\frac{2 a \sin ^2f_h}{r_h}+\frac{b \sin ^4f_h}{r_h^3}\right),\\
J=&m_{\pi }^2 r_h^4 \sin f_h+4 b \sin ^3f_h \cos f_h+2 a r_h^2 \sin 2 f_h,\\
H=&a r_h^4+2 b r_h^2 \sin ^2f_h+c \sin ^4f_h.\\
\end{split}
\ee
\begin{figure}[hbt]
\lbfig{sigmarh}
\begin{center}
\includegraphics[height=.36\textheight, trim = 60 20 80 50, clip = true]{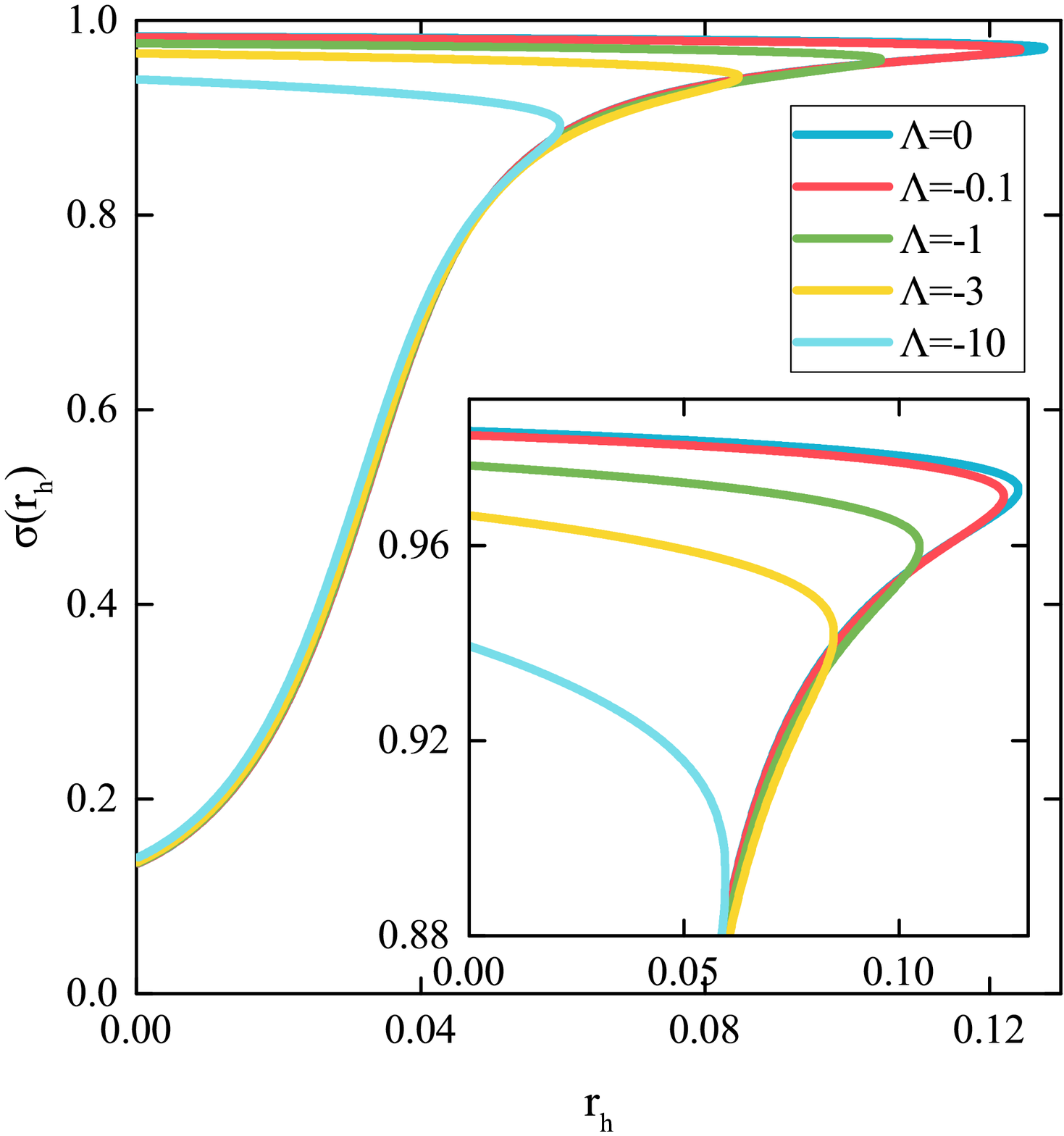}
\includegraphics[height=.36\textheight, trim = 60 20 80 50, clip = true]{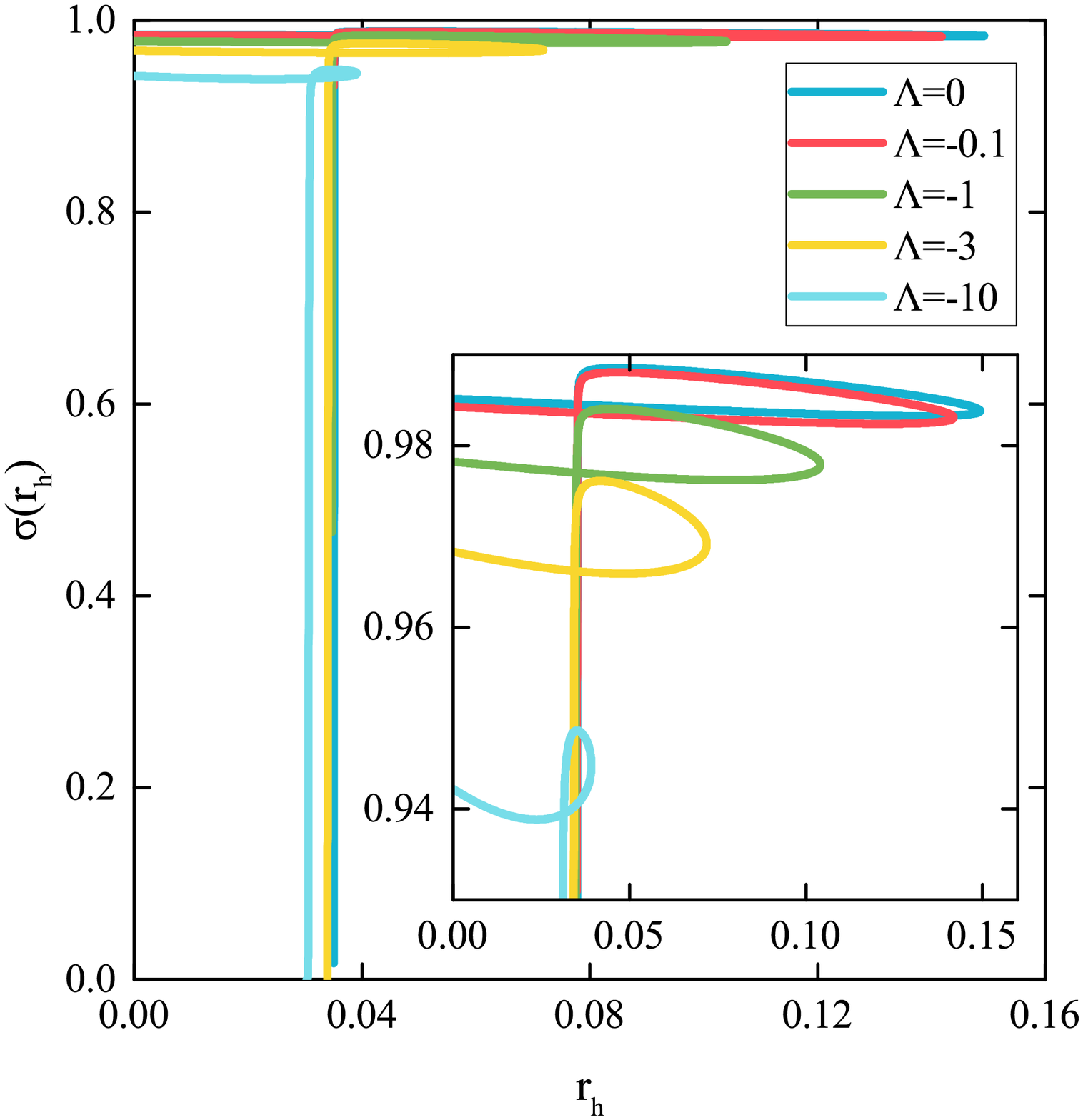}
\end{center}
\caption{\small The values of the metric function $\sigma(r_h)$ at the event horizon of the
$c=0$ submodel (left plot) and
the general  model at $c=1$ (right plot)
 are shown as functions of the
horizon radius $r_h$ at  $\alpha=0.05$ and some set of values of the cosmological constant $\Lambda$.}
\end{figure}

Here $f_h$ and $\sigma_h$ are the values of the Skyrme field profile function and the metric
function $\sigma$ at the event horizon, respectively. Note that it follows from the asymptotic expansion of the metric function
$N(r)$ that increasing the value of $|\Lambda|$ yields qualitatively the same effects as
increasing the value of the effective gravitational constant $\alpha$ \cite{Shiiki:2005aq}.

So, we have to impose the following set of the boundary conditions
on the event horizon and on the AdS asymptotic
\be
\label{bcondbh}
N\left(r_h\right)=0,\quad M^\prime(\infty)=0, \quad f(\infty)=0, \quad \sigma(\infty)=1.
\ee
The input parameters in the numerical solution of the system of equations \re{eqs} are
$f_h$, $\sigma_h$, $r_h$, $\alpha$ and $\Lambda$.

\begin{figure}[hbt]
\lbfig{rhT}
\begin{center}
\includegraphics[height=.26\textheight, trim = 60 20 80 50, clip = true]{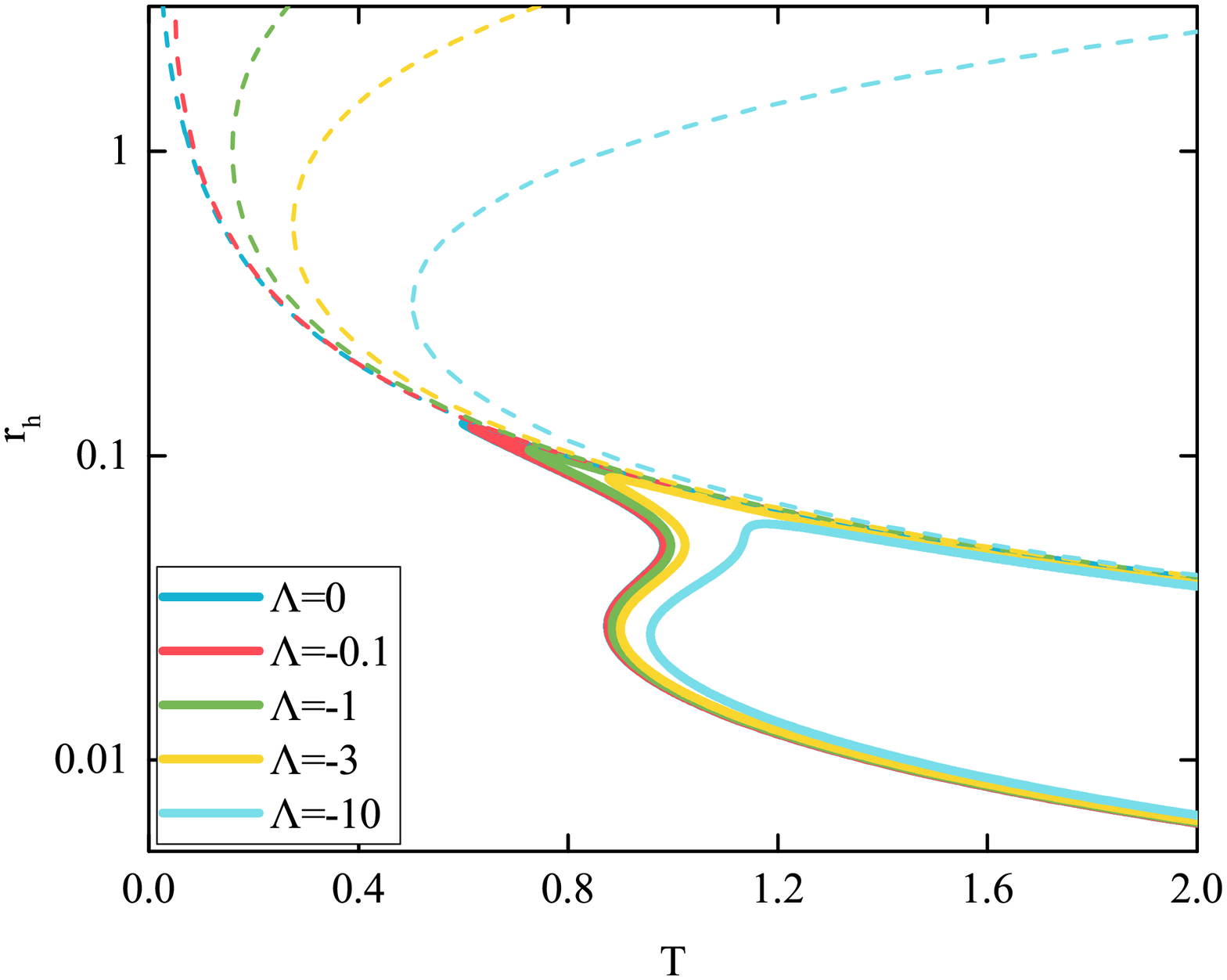}
\includegraphics[height=.26\textheight, trim = 60 20 80 50, clip = true]{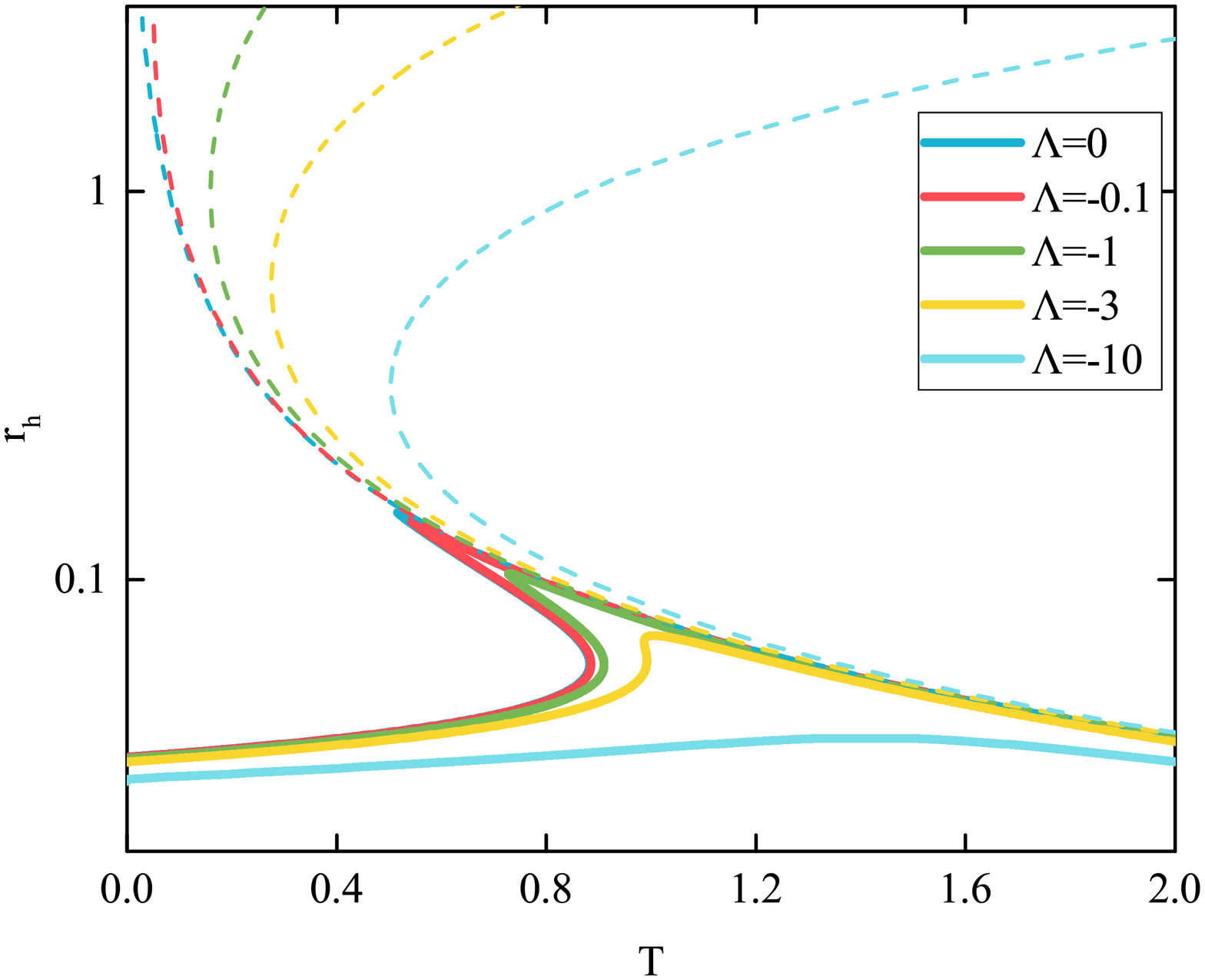}
\end{center}
\caption{\small
The Hawking temperature is shown as a function of the event horizon radius for several sets of solutions
of $c=0$ submodel (left) and $c=1$ general model (right)  at different values of
$\Lambda$ for  $\alpha=0.05$, $a=b=m_\pi=1$. The corresponding temperatures of the Schwarzschild-AdS black holes are
plotted as reference curves.
}
\end{figure}

\begin{figure}[hbt]
\lbfig{SM}
\begin{center}
\includegraphics[height=.26\textheight, trim = 60 20 80 50, clip = true]{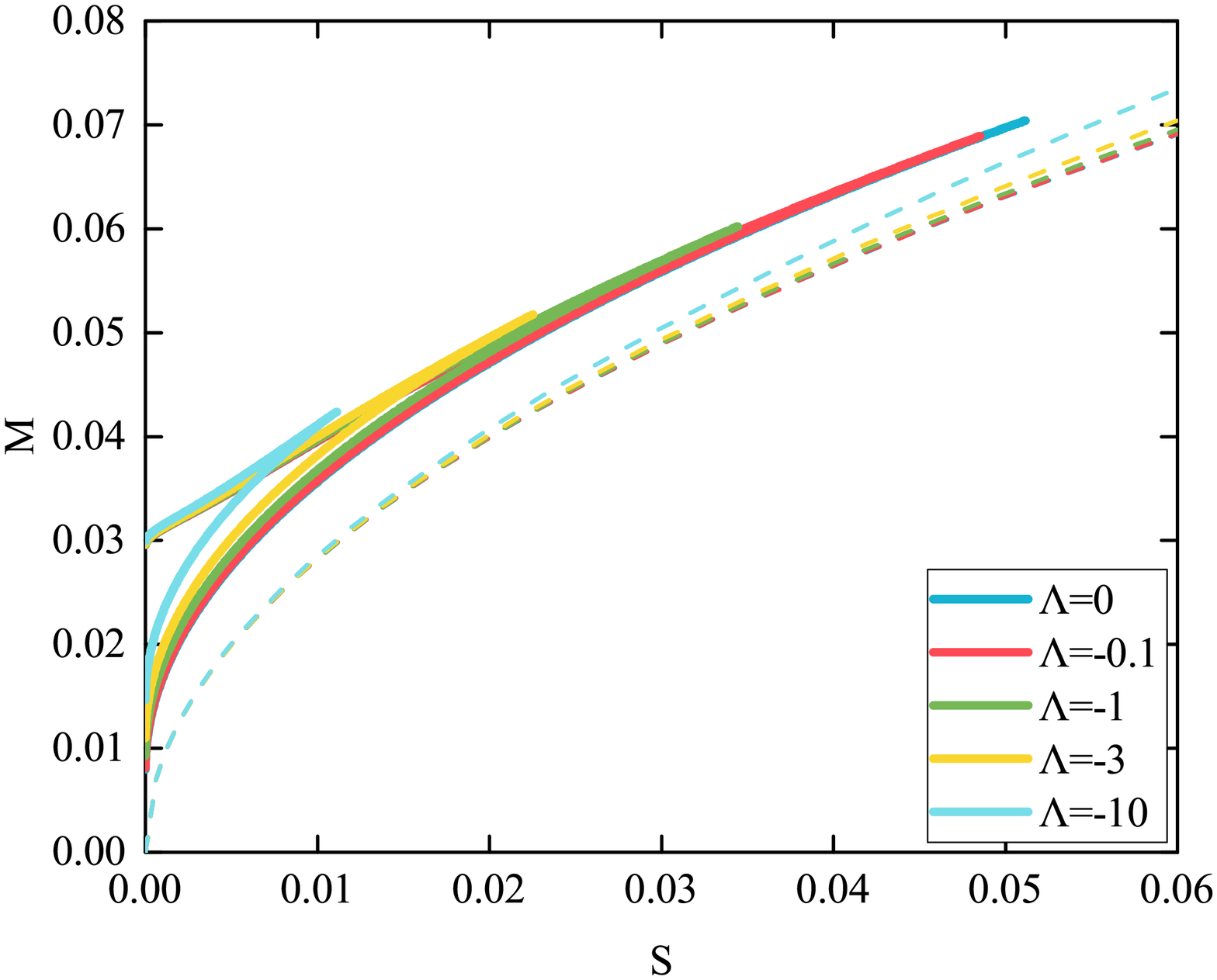}
\includegraphics[height=.26\textheight, trim = 60 20 80 50, clip = true]{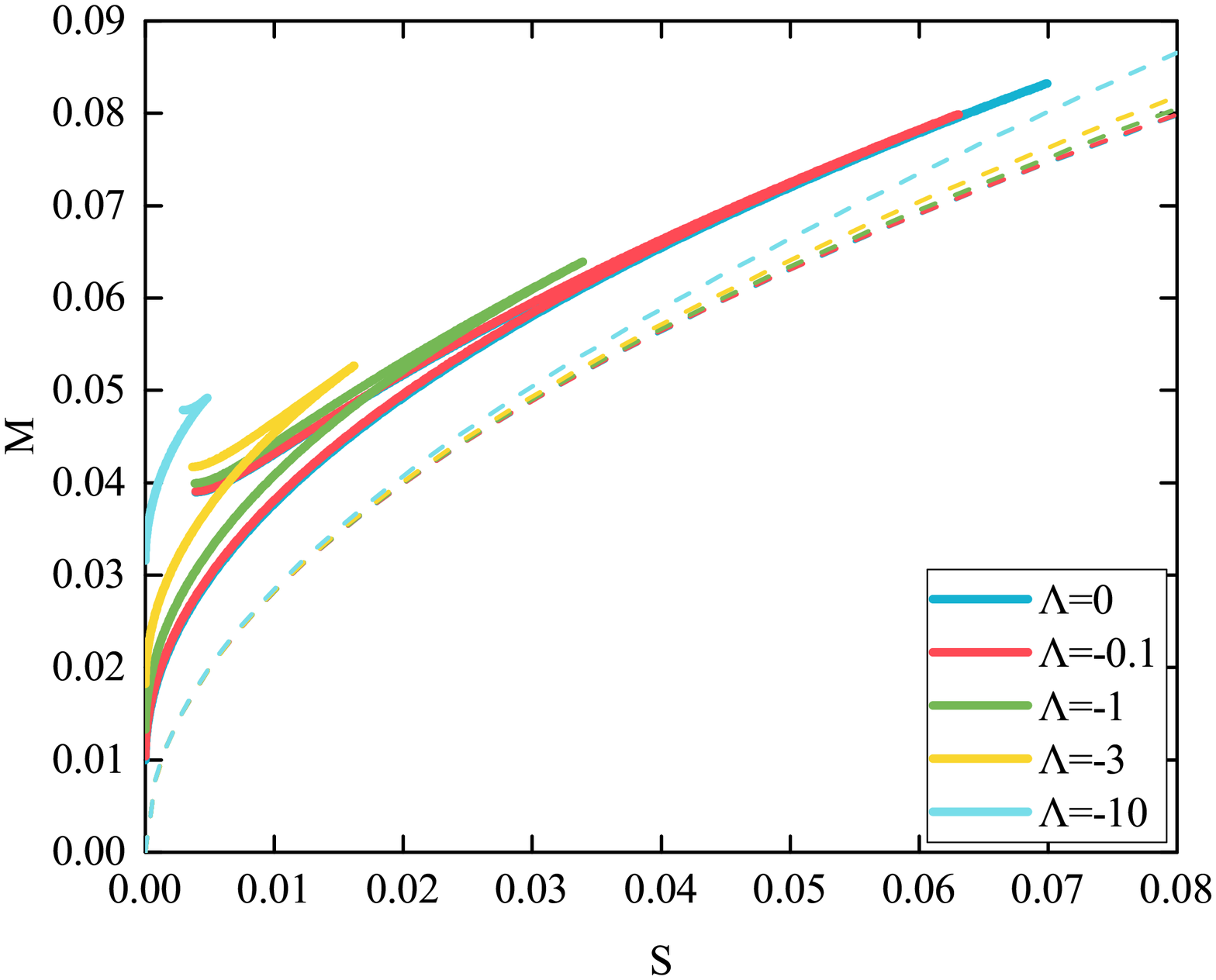}
\end{center}
\caption{\small
The mass-entropy curves for the static black holes in the general Einstein-Skyrme model are plotted
for several sets of solutions
of $c=0$ submodel (left) and $c=1$ general model (right)  at different values of
$\Lambda$ for  $\alpha=0.05$, $a=b=m_\pi=1$. The corresponding dependencies of the Schwarzschild-AdS black holes are
plotted as reference curves.
}
\end{figure}

\begin{figure}[hbt]
\lbfig{clim}
\begin{center}
\includegraphics[height=.28\textheight, trim = 50 20 80 50, clip = true]{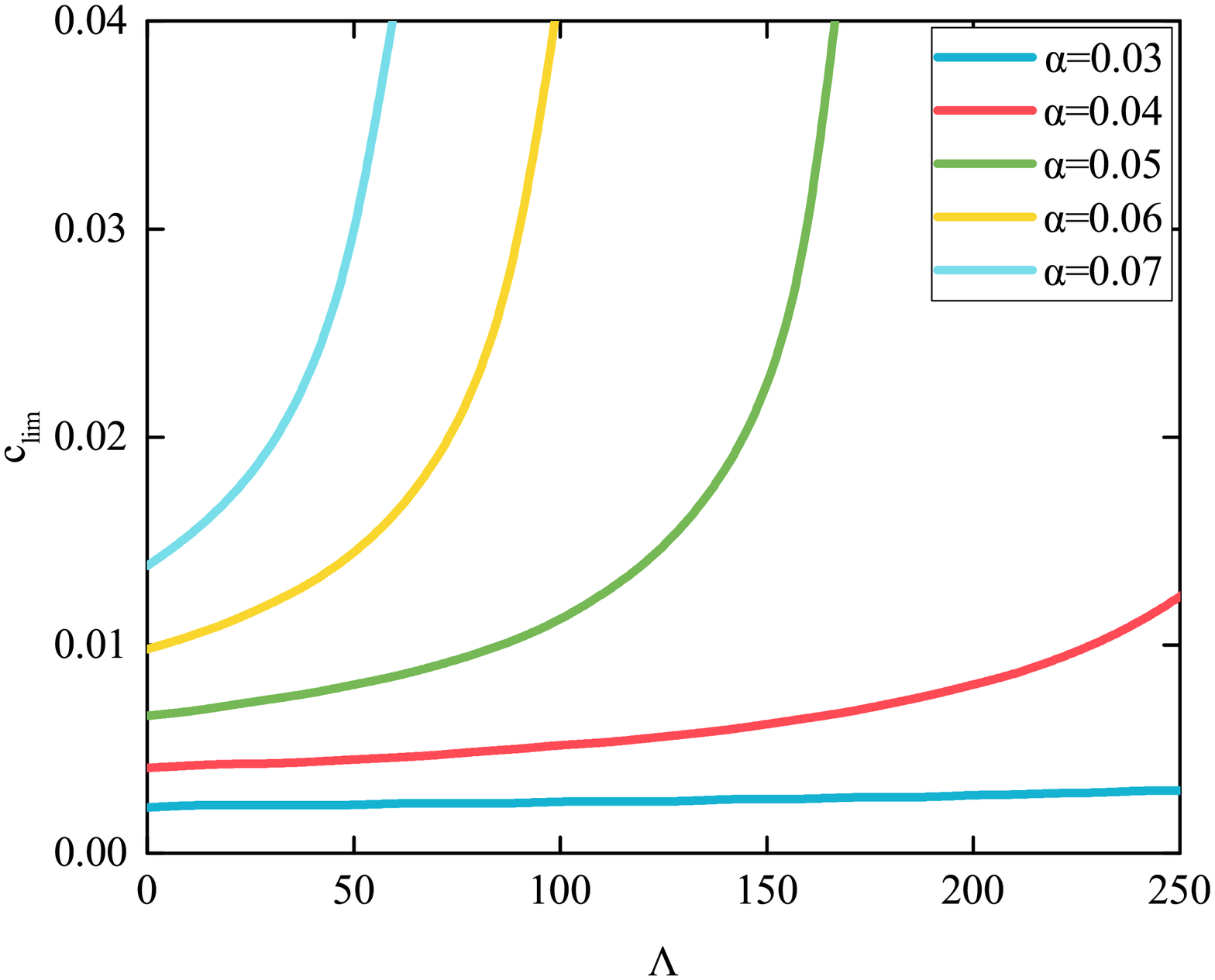}
\end{center}
\caption{\small
The critical value of the coupling $c$, below which the lower and upper
branches of the hairy black holes are closed, is shown as a function of the cosmological constant
$\Lambda$ for some set of values of $\alpha$.
}
\end{figure}

\subsection{Numerical results}

The properties of the spherically symetric black holes with general Skyrme hair in the asymptotically flat case were
considered before \cite{Adam:2016vzf,Gudnason:2016kuu}, now we can extend the consideration to the asymptotically AdS spacetime.
Indeed, we found that all known $\Lambda=0$ black hole solutions of the general Einstein-Skyrme model possess generalizations
with AdS asymptotics. For a fixed nonzero value
of the effective coupling $\alpha$, the lower in energy $r_h$-branch of solutions smoothly arises from the
corresponding regular self-gravitating configuration. This configuration can be viewed as a small
AdS-Schwarzschild black hole, embedded into the Skyrmion \cite{Ashtekar}.

\begin{figure}[hbt]
\lbfig{rhcrit}
\begin{center}
\includegraphics[height=.28\textheight, trim = 50 20 80 50, clip = true]{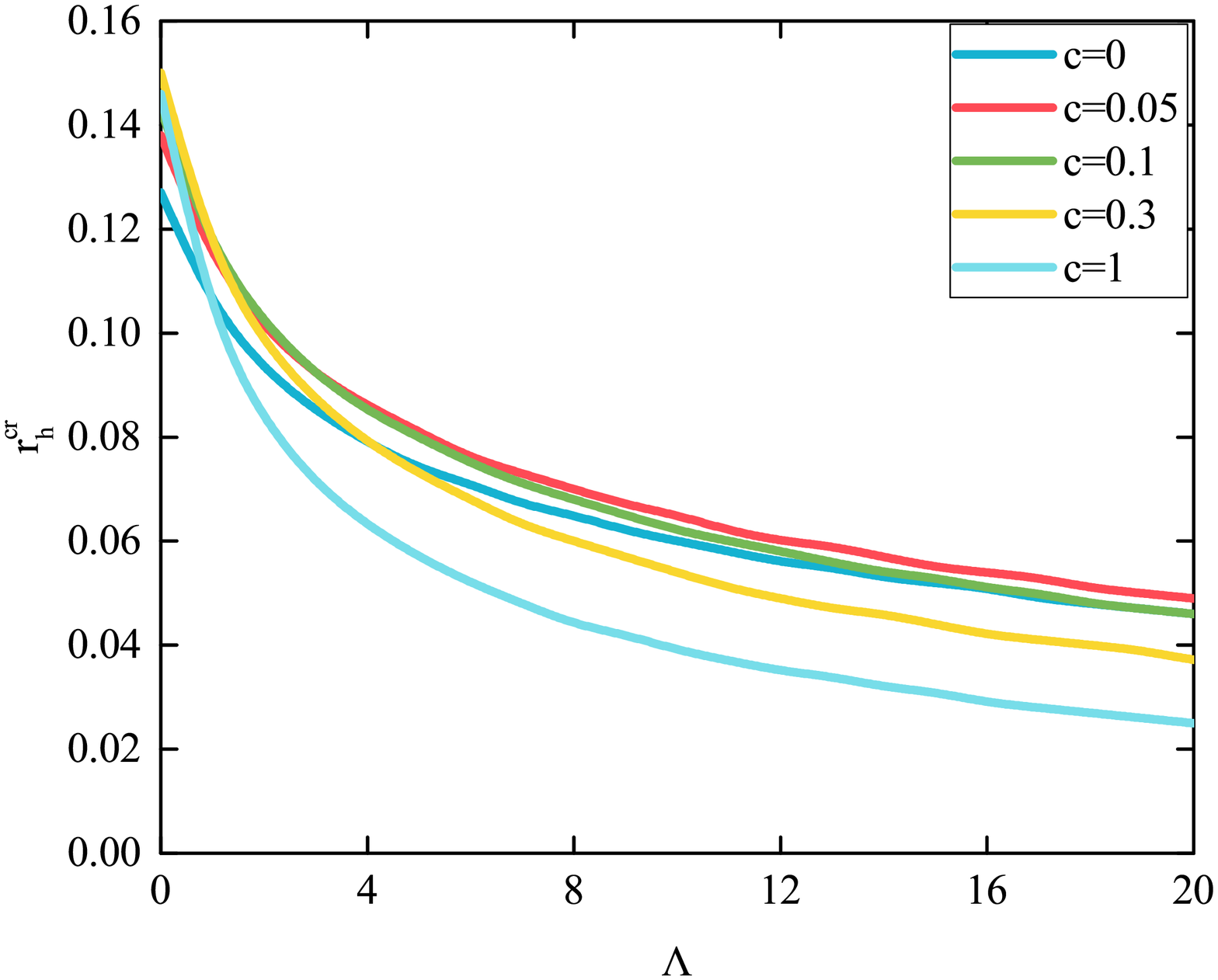}
\end{center}
\caption{\small
The critical maximal value of the event horizon radius $r_h^{cr}$ is shown as a function of the cosmological constant $\Lambda$
for some set of values of $c$ for $\alpha=0.05$ and $a=b=m_\pi=1$.
}
\end{figure}

\begin{figure}[hbt]
\lbfig{fTlambda}
\begin{center}
\includegraphics[height=.26\textheight, trim = 60 20 80 50, clip = true]{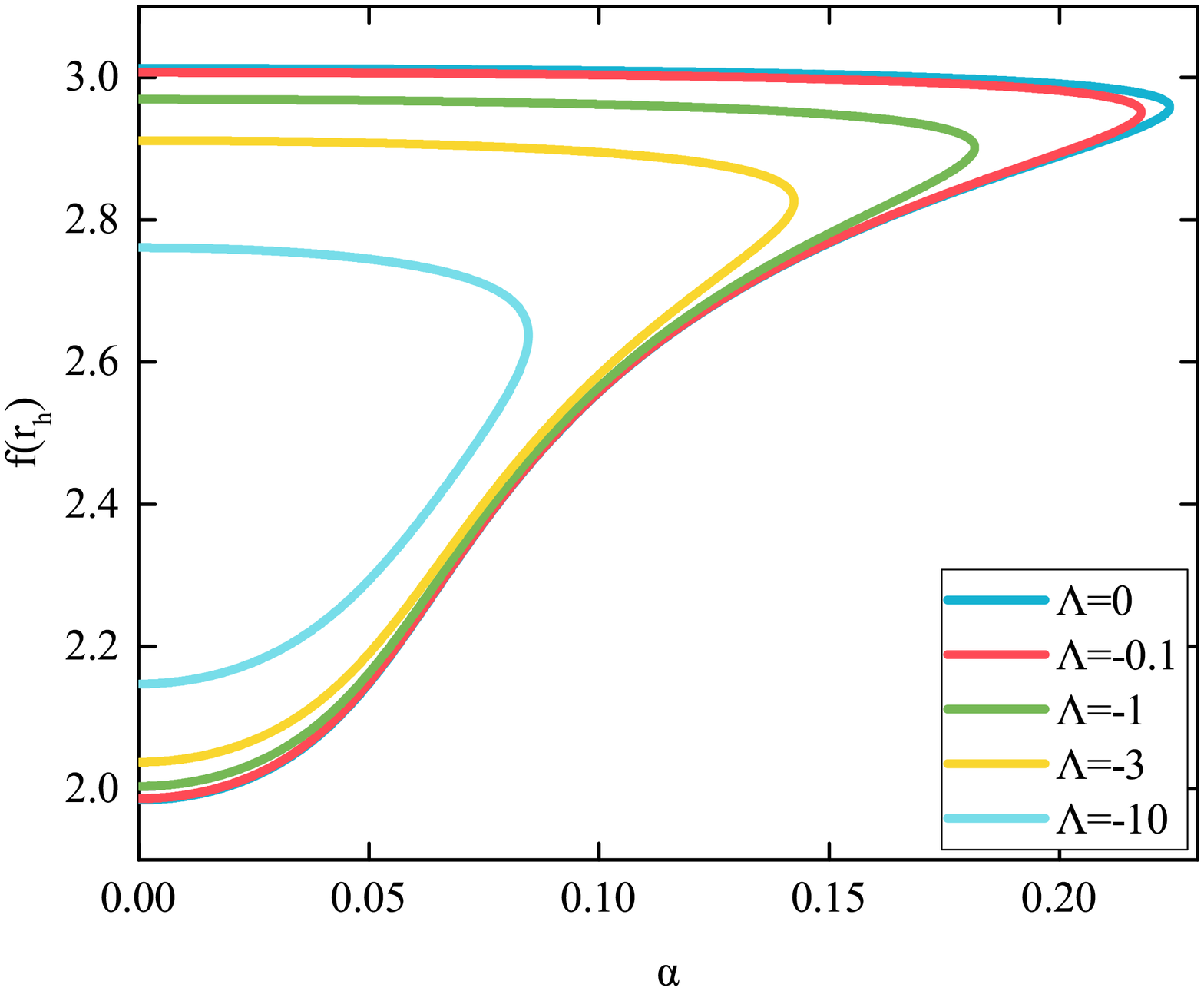}
\includegraphics[height=.26\textheight, trim = 60 20 80 50, clip = true]{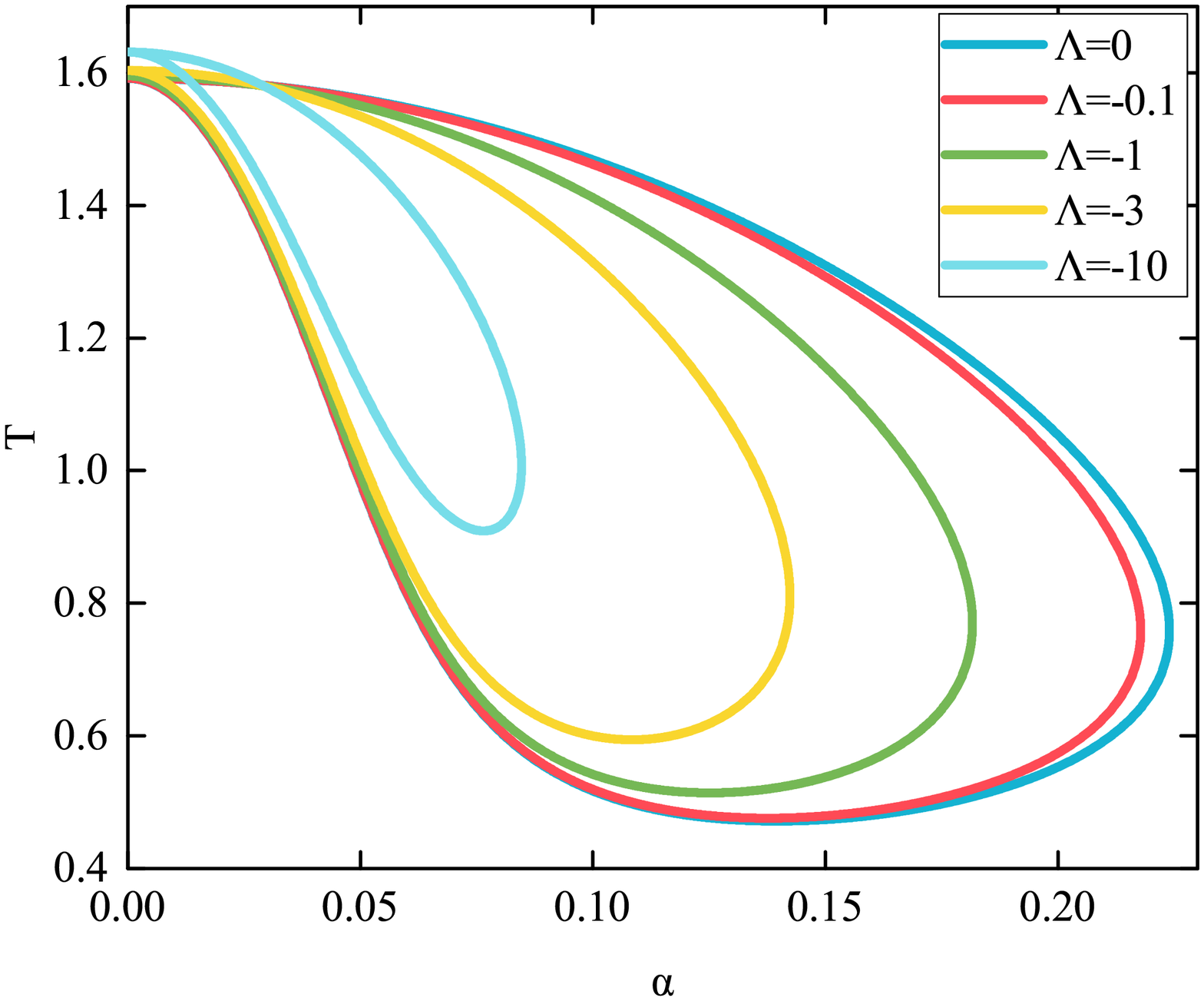}
\includegraphics[height=.26\textheight, trim = 60 20 80 50, clip = true]{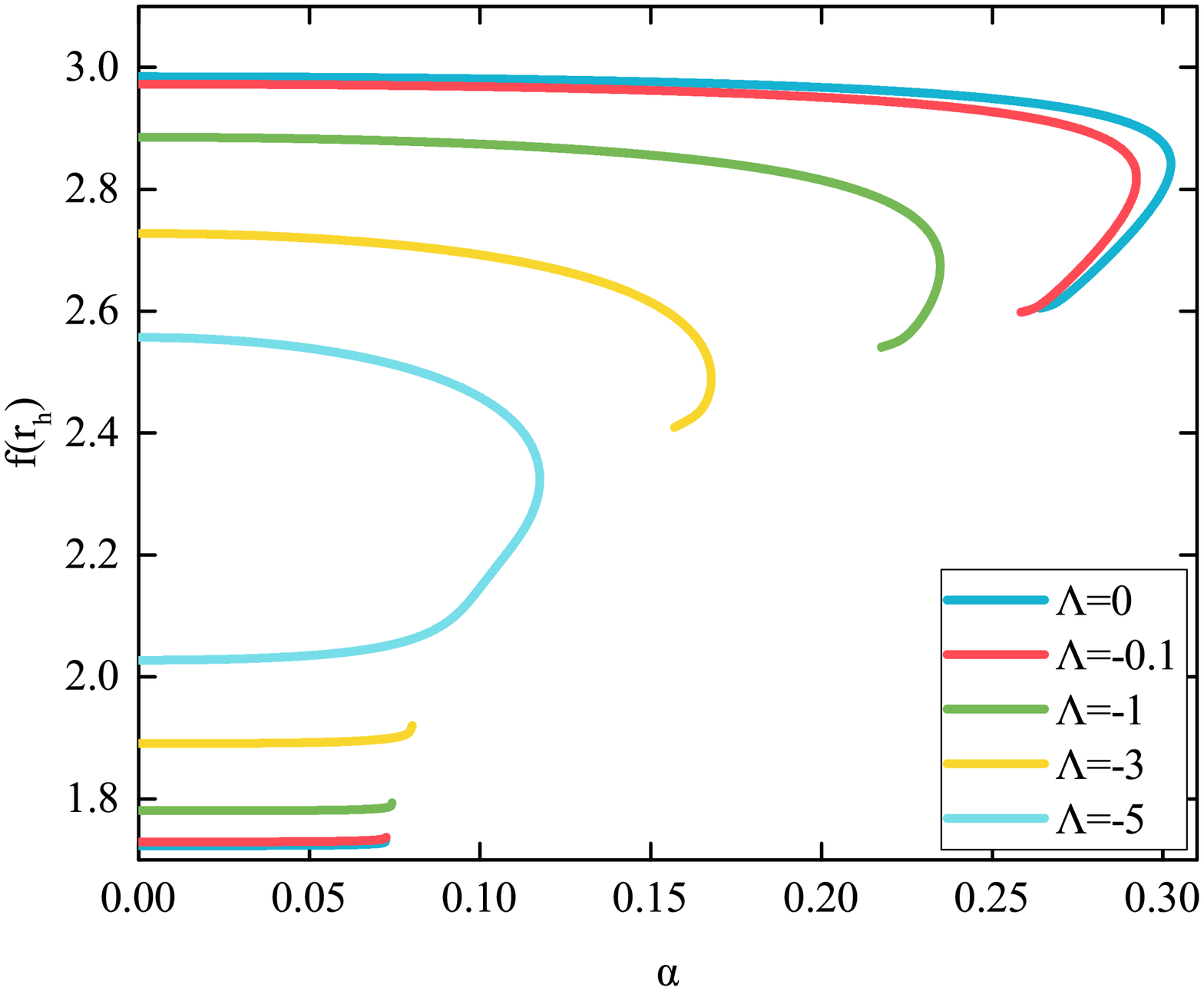}
\includegraphics[height=.26\textheight, trim = 60 20 80 50, clip = true]{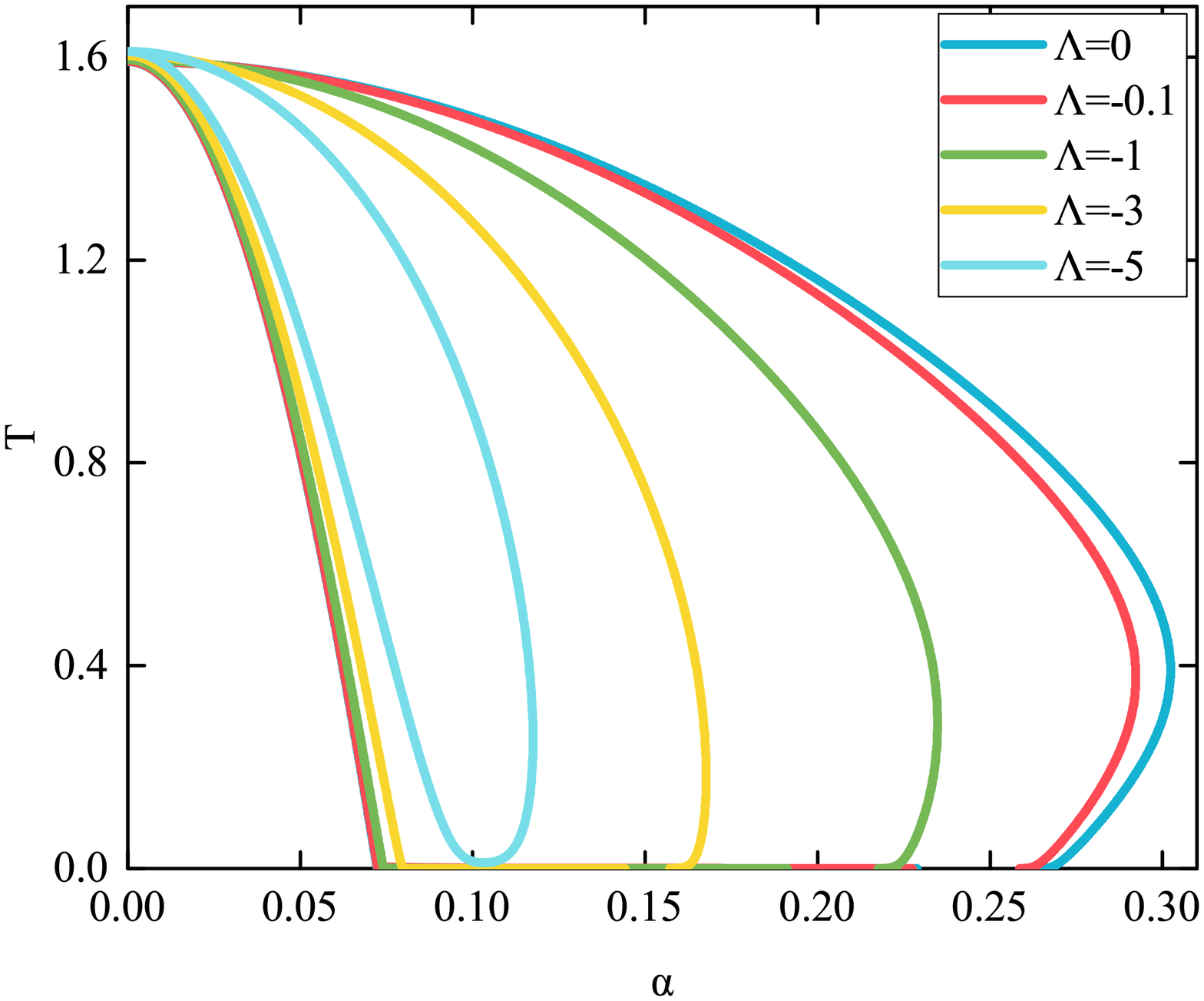}
\end{center}
\caption{\small
The values of the Skyrme field profile function at the event horizon, $f(r_h)$, (left column )
and the Hawking temperature of the Skyrme black hole (right column) are plotted as functions of
the effective gravitational coupling constant
$\alpha$ for $c=0$ (upper row) and $c=1$ (lower row), $r_h=0.05$ and some set of
values of cosmological constant $\Lambda$.
}
\end{figure}

\begin{figure}[hbt]
\lbfig{fMa}
\begin{center}
\includegraphics[height=.26\textheight, trim = 60 20 80 50, clip = true]{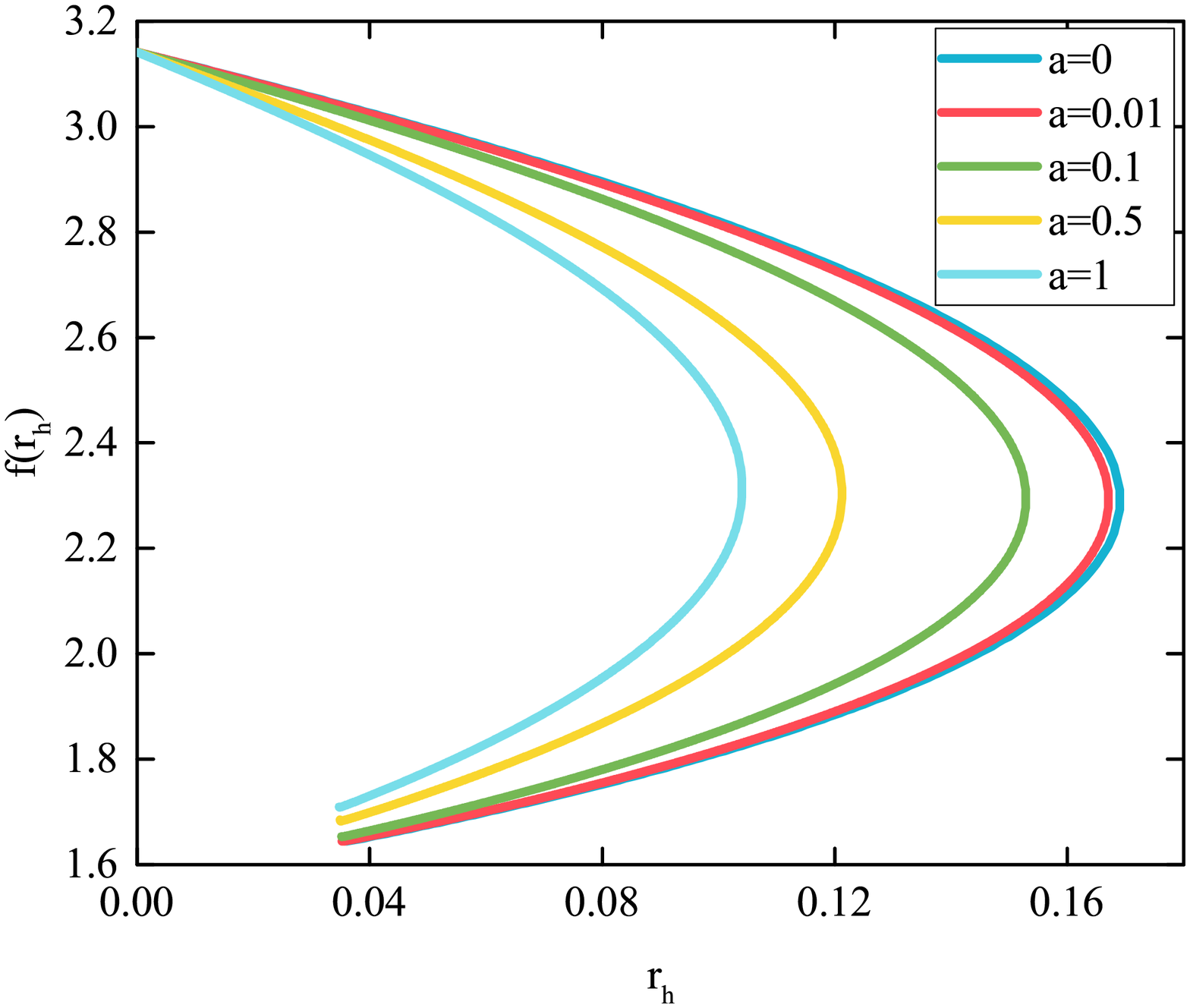}
\includegraphics[height=.26\textheight, trim = 60 20 80 50, clip = true]{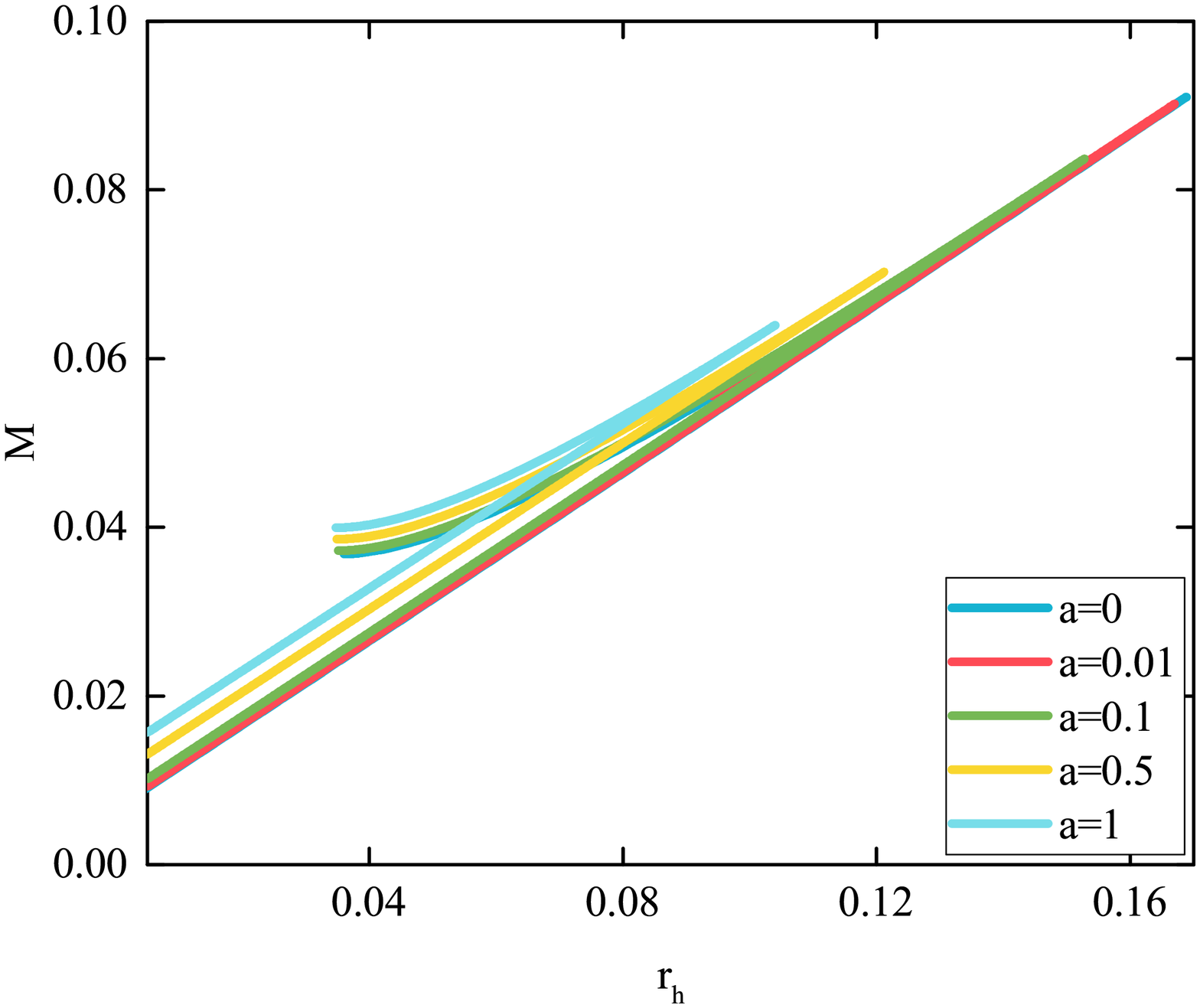}
\includegraphics[height=.26\textheight, trim = 60 20 80 50, clip = true]{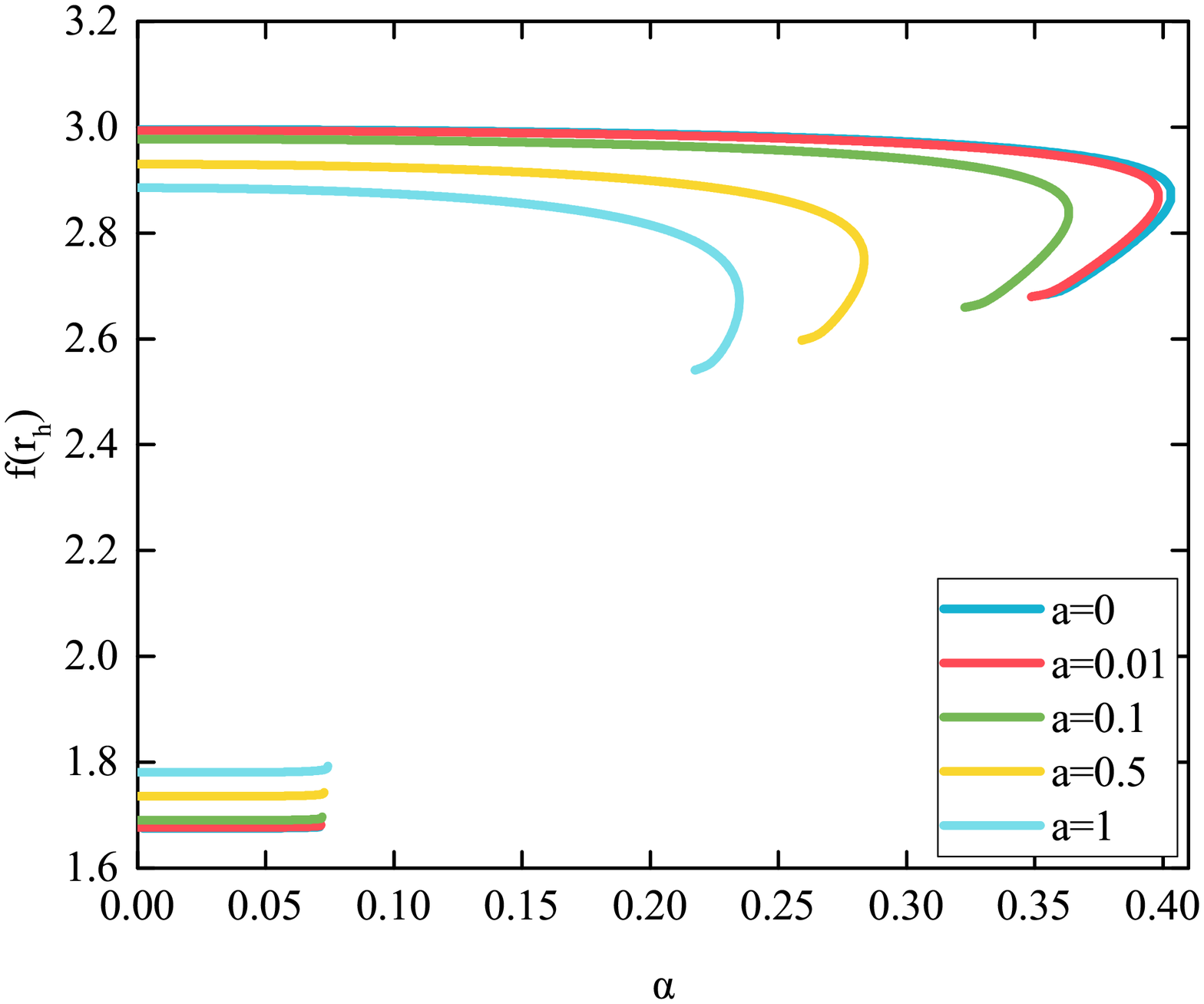}
\includegraphics[height=.26\textheight, trim = 60 20 80 50, clip = true]{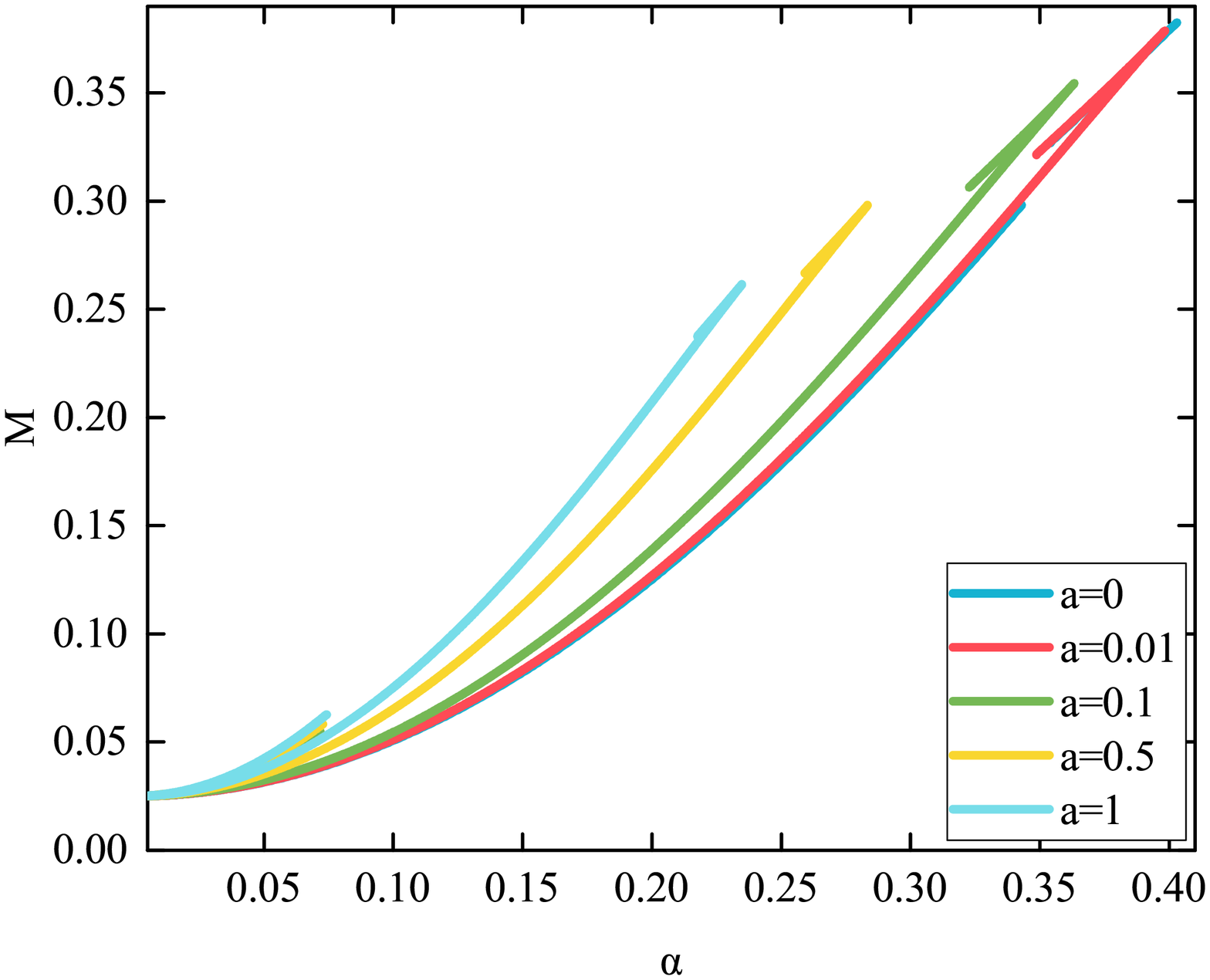}
\end{center}
\caption{\small
The value of the Skyrme field profile function at the event horizon, $f(r_h)$, (left column) and the ADM mass
of the solution (right column) are plotted as functions of the event horizon radius $r_h$ for $\alpha=0.05$ (upper row),
and the effective gravitational coupling constant $\alpha$ for $r_h=0.05$ (lower row)
for some set of values of $a$ at $b=c=m_\pi=1$ and $\Lambda=-1$.
}
\end{figure}

In Fig.~\ref{fMlambda} we present some results of the numerical integration for two sets of solutions with $c=0$ and
$c=1$ at $\alpha=0.05$ and $a=b=m_\pi=1$. The control parameter of the numerics is $r_h$,
which fixes the size of the event horizon.

We observe that for any values of $c$, $\Lambda$ and, for allowed on the lower regular branch
value of the effective gravitational constant $\alpha$, the corresponding
branch of black hole solutions always emerges  from
the corresponding regular self-gravitating Skyrmion with a vanishing horizon area $r_h\to 0$ and a nonzero mass.
As expected, in general there is not much
difference between the pattern of evolutions of the solutions of the $c=0$ submodel and the $c=1$ general model along this
lower in mass branch. Both the values of the Skyrme
field profile function $f(r_h)$ and the metric function  $\sigma(r_h)$ on the event horizon are
monotonically decreasing as the horizon radius $r_h$ increases, see Fig.~\ref{fMlambda}. The mass of the
configurations is increasing along this branch, see  Fig.~\ref{fMlambda}, while the value of the metric function $\sigma_h$ at the
horizon slowly decreases, see Fig.~\ref{sigmarh}. As seen in Fig.~\ref{rhT}, on the lower branch  the Hawking temperature
decreases  as $r_h$ grows.

The lower branch always terminates at come critical maximal value of the horizon radius $r_h^{cr}$. There it
bifurcates with a secondary, upper (higher-mass) branch, which extends
backward in $r_h$. The upper branch solutions have higher entropy than the solutions on the
lower branch.

The curves, which shows the dependency of the mass of the configuration on the horizon radius,
form a typical cusp at the critical point, as exhibited in Fig.~\ref{fMa}, right plots.
As expected, the value of the critical horizon radius $r_h^{cr}$ decreases as absolute value of the
cosmological constant $\Lambda$ grows, see  Fig.~\ref{rhcrit}.

Remarkably, the pattern of evolution of the hairy black holes along the upper branch  critically depends on the
value of the parameter $c$ \cite{Adam:2016vzf,Gudnason:2016kuu}, although the bifurcation into
two $r_h$ branches continues to hold for any value of $c$. In the Einstein-Skyrme submodel with $c= 0$
the upper branch always extends backwards to the limit $r_h=0$ for any values of $\Lambda$ and for any
$\alpha$, see  Figs.~\ref{fMlambda},\ref{sigmarh}. As $|\Lambda|$ grows, this branch becomes shorter, however
it always approach the limiting $r_h=0$ solution on the corresponding upper branch of the regular self-gravitating
Skyrmions in asymptotically AdS spacetime. This case was considered before in \cite{Shiiki:2005aq}, our results nicely agree
with observations reported in this paper. As one can see in Figs.~\ref{fMa}, along the upper branch the value of the Skyrme field
profile function on the event horizon initially continues to decrease as $r_h$ decreases, it approaches some minimal value and then
rapidly increases up to $f(r_h)=\pi$ in the limit $r_h\to 0$. The minimal value of the function $f(r_h)$ slightly increases as
$|\Lambda|$ increases. The value of the second metric function on the horizon $\sigma(r_h)$ on the upper branch is rapidly decreasing,
as expected, see Fig.~\ref{sigmarh}, left plot.

Interestingly, the Hawking temperature of these  hairy black holes
exhibits non-monotonical behavior, it starts to increase as $r_h$ decreasing along this branch, then on some subinterval of
values of $r_h$ it decreases again, and finally it rapidly grows as $r_h$ continues to decrease, see Fig.~\ref{rhT}, left plot.
Thus, for some interval of values of the temperature, there are three solutions with different masses at a given $T$. For relatively
small values of $|\Lambda|$ the absolute minimum of the temperature corresponds to the maximal radius of the horizon $r_h^{cr}$.
As $|\Lambda|$ increases, the second minimum of the temperature becomes the absolute minimum.

As we can see, for a fixed value of $r_h$
the Hawking temperature of the Skyrme black hole decreases
as  $|\Lambda|$ increases, although for the Schwarzschild-Anti-de Sitter configuration it increases \re{TAdS}.
To explain this observation we notice that, as in follows from \re{hawking}, the
temperature is proportional to the product of values of the metric function $\sigma$ and the derivative of the
second metric function $N(r)$ at the event horizon.
While $\sigma_h$ decreases with increasing of $\Lambda$, cf. Fig.~\ref{sigmarh},
the derivative of the metric function $N$ increases, see \re{horexp2}. So, the Hawking temperature
$T_H$ decreases because  $\sigma_h$ decreases faster than $N'(r_h)$
increases as $|\Lambda|$ grows.

\begin{figure}[hbt]
\lbfig{fMb}
\begin{center}
\includegraphics[height=.26\textheight, trim = 60 20 80 50, clip = true]{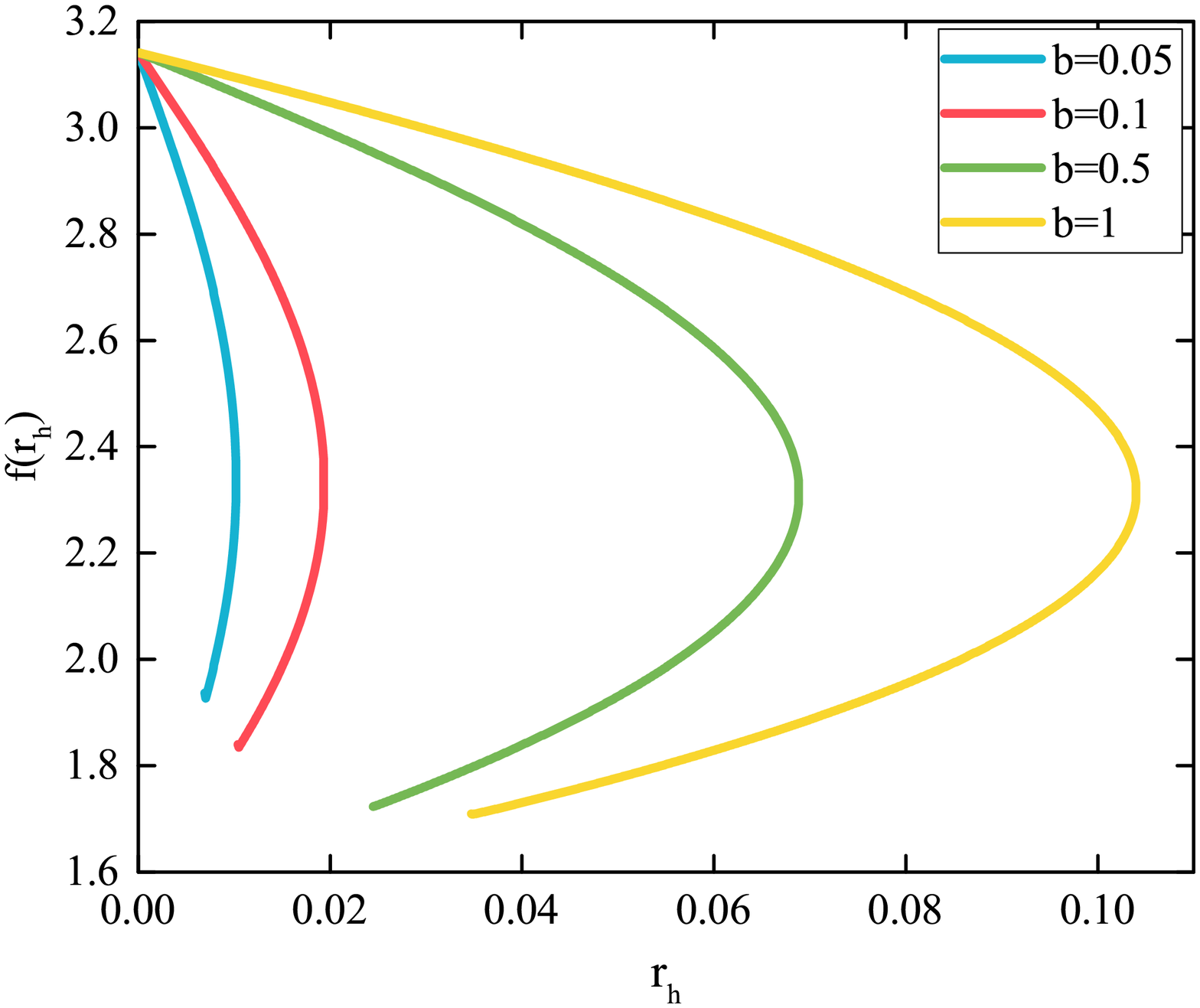}
\includegraphics[height=.26\textheight, trim = 60 20 80 50, clip = true]{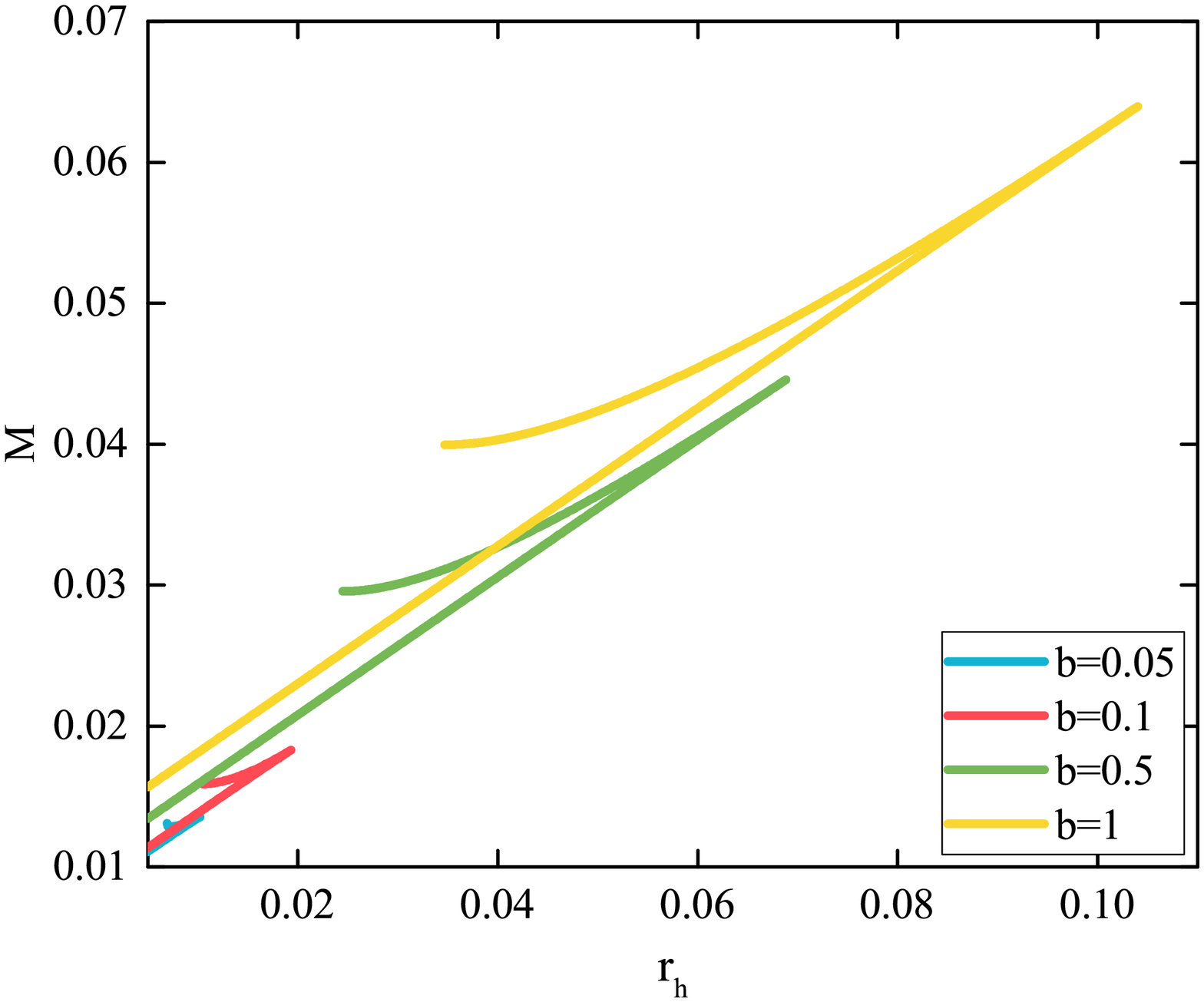}
\includegraphics[height=.26\textheight, trim = 60 20 80 50, clip = true]{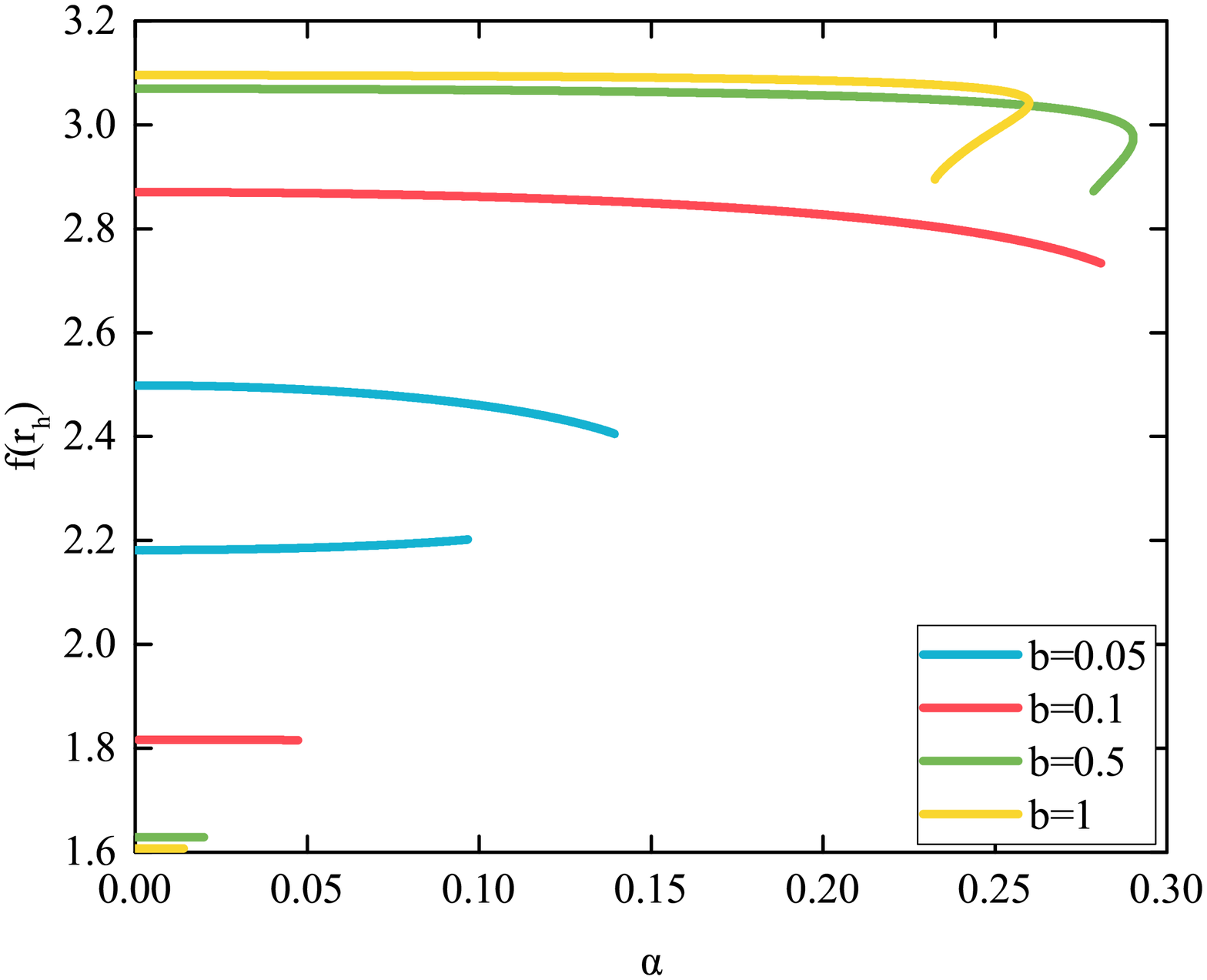}
\includegraphics[height=.26\textheight, trim = 60 20 80 50, clip = true]{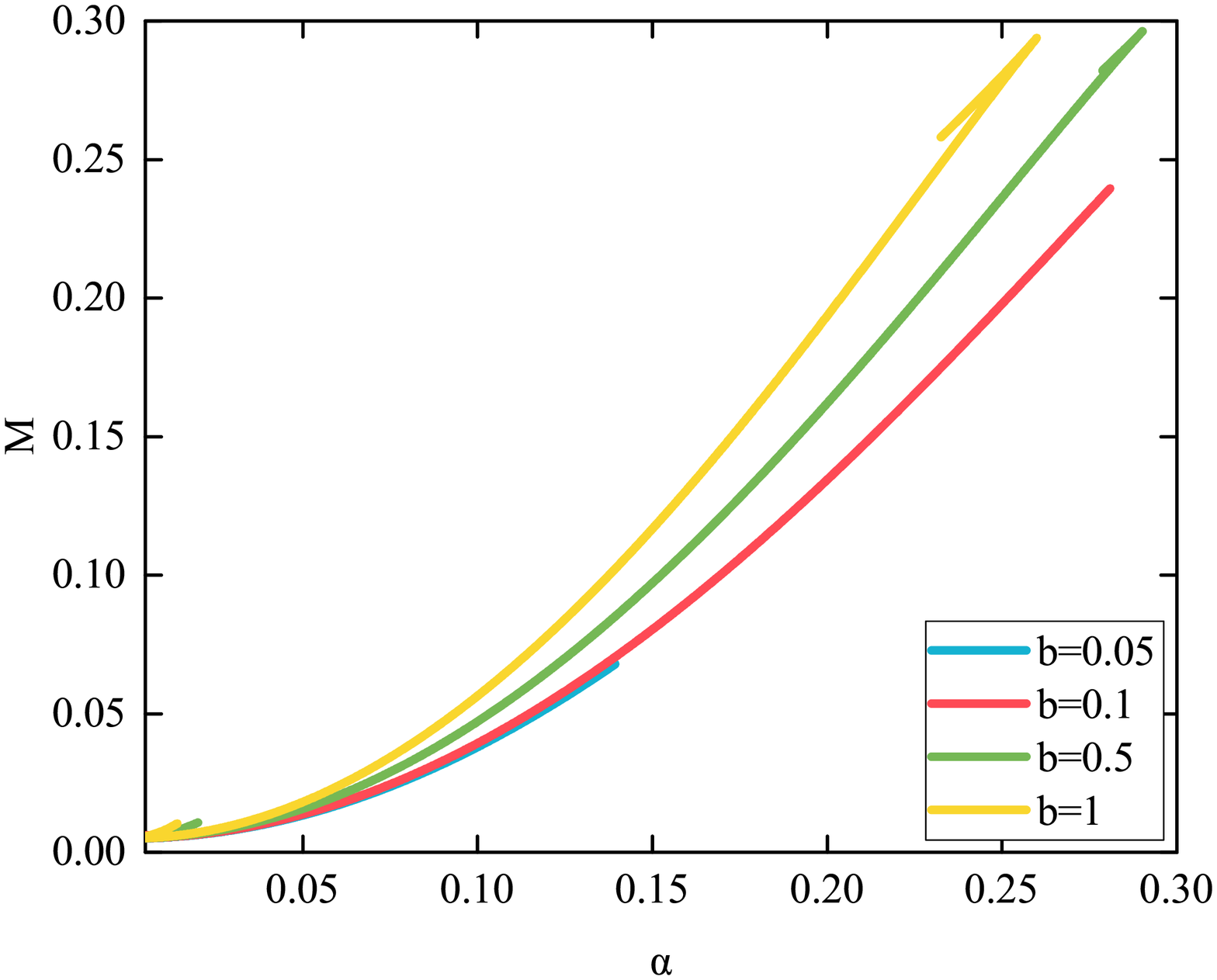}
\end{center}
\caption{\small
The value of the Skyrme field profile function at the event horizon, $f(r_h)$, (left column)
and the ADM mass of the solutions (right column) as functions of the event horizon radius $r_h$ at $\alpha=0.05$ (upper row)
and the effective gravitational coupling constant $\alpha$ at $r_h=0.01$ (lower row) for some values of $b$, $a=c=m_\pi=1$,
and $\Lambda=-1$.
}
\end{figure}

It may be interesting to consider the thermodynamic properties of the hairy black holes in the general Einstein-Skyrme model in AdS spacetime
and compare it with the pattern observed in other related systems, see e.g.
\cite{Maeda:1993ap,Winstanley:1998sn,Radu:2011uj,Kichakova:2015lza}.
In Fig.~\ref{SM} we present the mass-entropy relation of the solutions, together with the corresponding curves of the
vacuum Schwarzschild-AdS black holes.
First, we observe that the two-branch structure exist both for the generalized $c=1$ Einstein-Skyrme model and for the $c=0$ submodel.
The upper and lower branches described above, correspond to the high- and low-entropy branches, respectively \cite{Shiiki:2005aq}.
The lower entropy $r_h$-branch always originates from the corresponding regular self-gravitating configuration,
the mass and the entropy of the hairy black hole are monotonously increasing along this branch. The curve $T(r_h)$ is just below
the corresponding lower branch of the Schwarzschild-AdS solution, see Fig.~\ref{rhT}.

The entropy and the mass are
maximal at the critical point, where the curve has a cusp merging the second, higher entropy branch. This branch is thermodynamically
unstable, as in the case of other hairy black holes \cite{Maeda:1993ap}. The mass and the values of the fields on the event horizon
$f(r_h)$, $\sigma(r_h)$ are monotonically decreasing as  $r_h$ is decreasing on this branch. As in the case of
the asymptotically flat space, for the general $c=1$ model the upper branch terminates at
the limiting (extremal) zero temperature solution, which still possess finite non-zero mass and entropy, see Fig.~\ref{SM},
there is no solutions  for values of the $r_h$ smaller than some critical value $r_h^{ex}$.
As $|\Lambda|$ increases, the maximal entropy and the mass of solutions decreases, the branches become shorter.
Evidently, the same effect yields increasing of the gravitational coupling $\alpha$ as other parameters of the model remain fixed.
The novel feature present in the case of the asymptotically AdS spacetime is that
the cosmological constant $\Lambda$ reduces the entropy.

As we have seen, the temperature of hairy black holes along the upper branch depends on $r_h$ nonmonotonically, it possess a
maximum at some point, see Fig.~\ref{rhT}. Note that, similar to the case of asymptotically flat spacetime,
the limiting behavior of the solution on this branch critically
depends on the value of coupling constant $c$ \cite{Adam:2016vzf,Gudnason:2016kuu}.
For $c\leq c_{lim}$, where $c_{lim}$ is some limiting value of $c$, the temperature, which is initially decreasing along this branch,
starts to increase rapidly as $r_h\to 0$. Thus, the $r_h$-branches are closed, in the sense that both the lower and the upper branches
are linked to the corresponding regular self-gravitating Skyrmions in the limit $r_h=0$ with infinite Hawking temperature.

In Fig.~\ref{profile}, we exhibit the profile functions
of the Skyrmion on the upper and lower branches for a few
values of the cosmological constant $\Lambda$. Clearly, one can see that the effect of variation of $\Lambda$ is much more significant
for the lower branch configurations.

While increasing the value of the cosmological constant $|\Lambda|$, and/or the effective gravitational coupling $\alpha$,
the critical value  $c_{lim}$ rapidly grows, see Fig.~\ref{clim}.
Physically,  we can expect that for any value of $\Lambda$ there
exists such a value of $\alpha$, for which both  $r_h$-branches are closed for any value of $c$ within range
$0\leq c \leq 1$.

Our numerical results further indicate that, for $c>c_{lim}$ the solutions on the unstable upper branch approach some
finite value at $r_h=r_h^{ex}$,
beyond which it cease to exist. At this point both the Hawking temperature of the solution and the value of the
metric function $\sigma$ at the event horizon, $\sigma(r_h)$, approach zero, see Figs.~\ref{rhT},\ref{sigmarh}.

\begin{figure}[hbt]
\lbfig{lambdamax}
\begin{center}
\includegraphics[height=.26\textheight, trim = 60 20 80 50, clip = true]{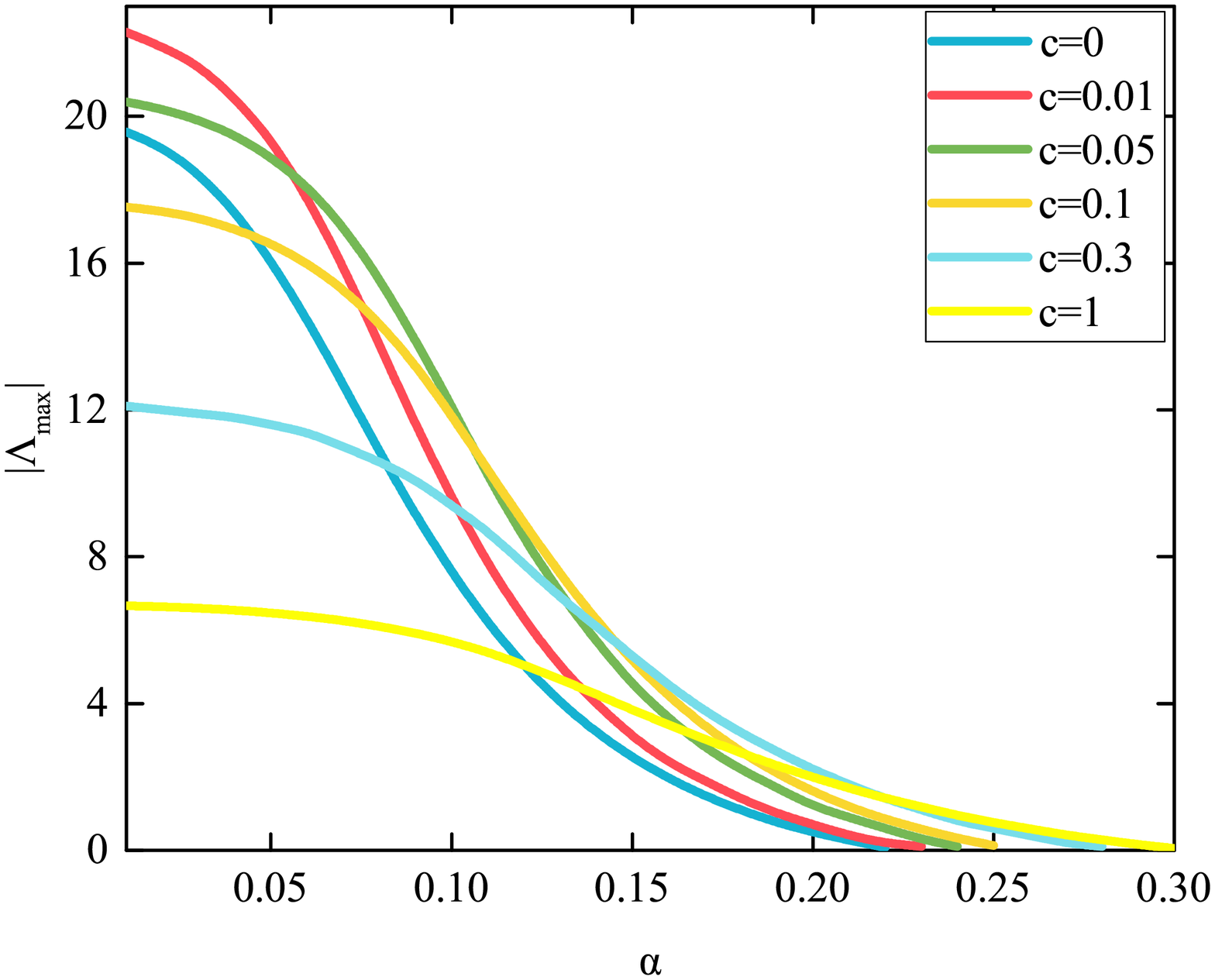}
\includegraphics[height=.26\textheight, trim = 60 20 80 50, clip = true]{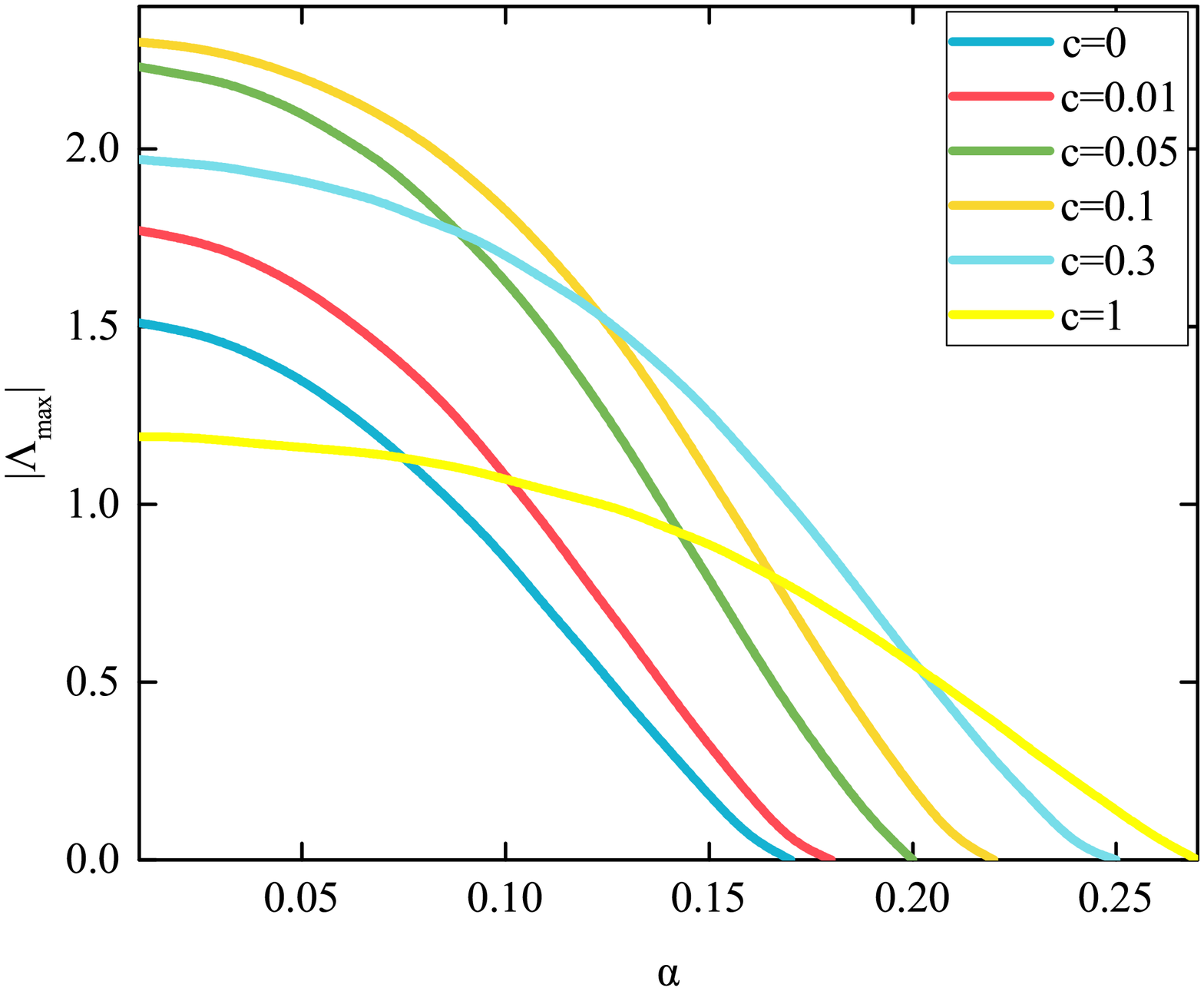}
\end{center}
\caption{\small
The maximal value of the cosmological constant $\Lambda_{max}$, for which hairy black holes may exist in a general Skyrme model,
is plotted as a function of the  effective gravitational coupling constant $\alpha$ for $r_h=0.05$ (left),
$r_h=0.1$ (right) and some set of values of $c$.
}
\end{figure}

It is worth to point out that the critical values  $r_h^{cr}$ and $r_h^{ex}$ typically depend on the coupling
constant $c$ and on $\Lambda$. Both of them are monotonically decreasing as $|\Lambda|$ increases, while $r_h^{ex}$
monotonically increases with increasing
of $c$. However the form of dependence of $r_h^{cr}$ on $c$ is a bit more complex, with a maximum for some value
of $r_h^{cr}$. In Fig.~\ref{rhcrit} we present the corresponding maximal values of the event
horizon radius as functions of $\Lambda$ for some set of values of $c$.

Let us also investigate the dependence of the hairy black hole solutions on the effective gravitational constant $\alpha$ for
a fixed value of the  horizon radius $r_h$. Here we also obtain two branches of solutions, linked to the corresponding configurations
on the lower and upper $r_h$ branches, respectively. These $\alpha$-branches bifurcate
at some maximal value, say  $\alpha^{cr}$, above which the gravitational interaction becomes
too strong for solutions to exist. When $r_h$ and $\Lambda$ are fixed, the mass of the configuration on the first branch
increases as $\alpha$ increases, see Fig.~\ref{fMa}.

As expected, the upper $\alpha$-branch solutions may exist along the entire branch only for $c\leq c_{lim}$. Physically, it
corresponds to the link to the limiting configuration, which represents the black hole with Bartnik-McKinnon hairs
\cite{Volkov:1989fi,Bizon:1990sr}. As $c>c_{lim}$, there are no solutions for the range of parameters
$\alpha_1^{ex}<\alpha<\alpha_2^{ex}$, where at the endpoints $\alpha_1^{ex}$ and $\alpha_2^{ex}$ both the Hawking temperature
and the value of the metric function $\sigma$ at the event horizon, approach zero.
While increasing the value of the cosmological constant $\Lambda$ the numerical analysis shows
that the endpoints $\alpha_1^{ex}$ and $\alpha_2^{ex}$ are coming closer and temperature of black hole increases, thus there is a critical
value of $\Lambda$ for which the singular solutions on the seconds branch cease to exist, see  Fig.~\ref{fTlambda}.

\begin{figure}[hbt]
\lbfig{profile}
\begin{center}
\includegraphics[height=.26\textheight, trim = 60 20 80 50, clip = true]{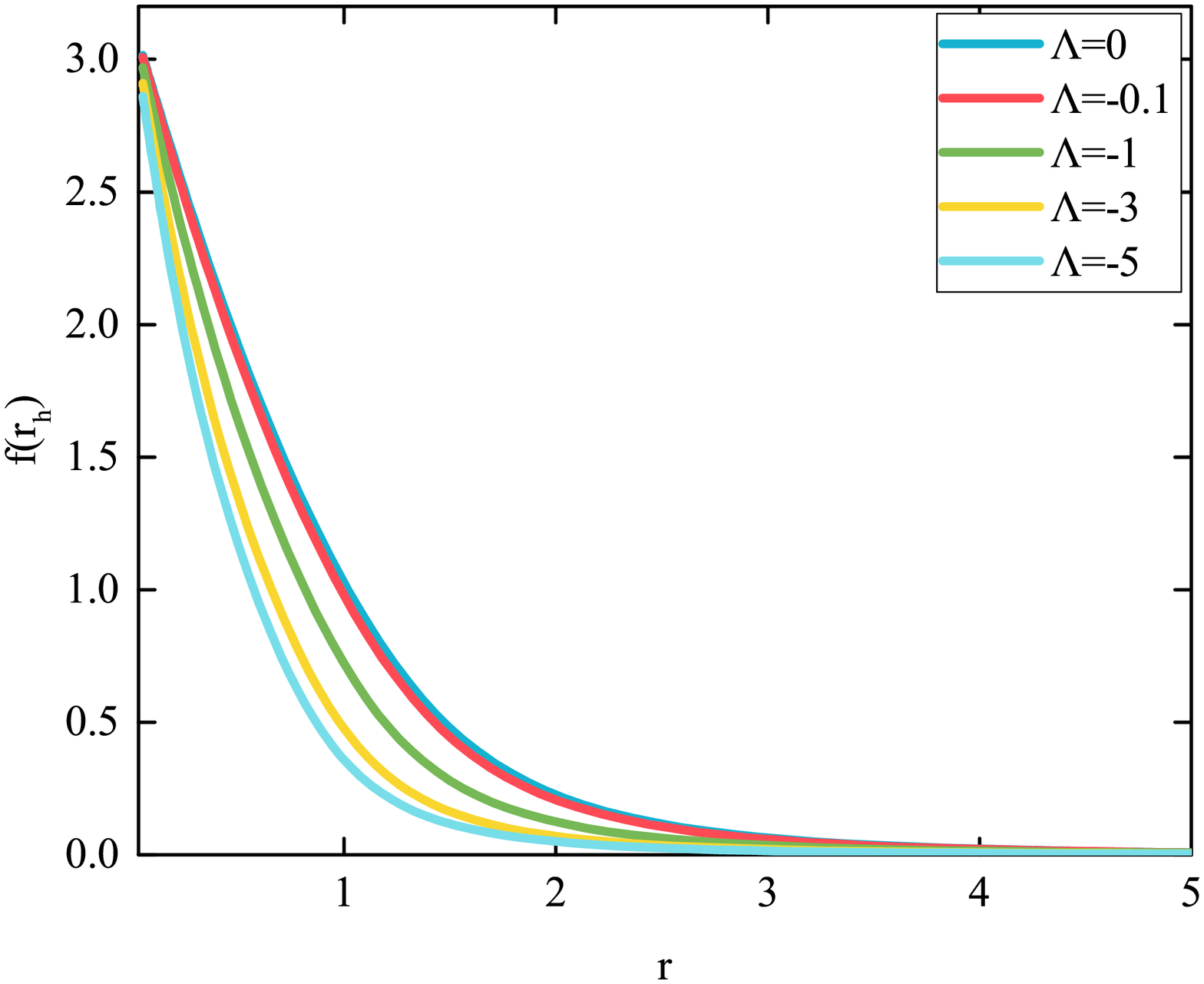}
\includegraphics[height=.26\textheight, trim = 60 20 80 50, clip = true]{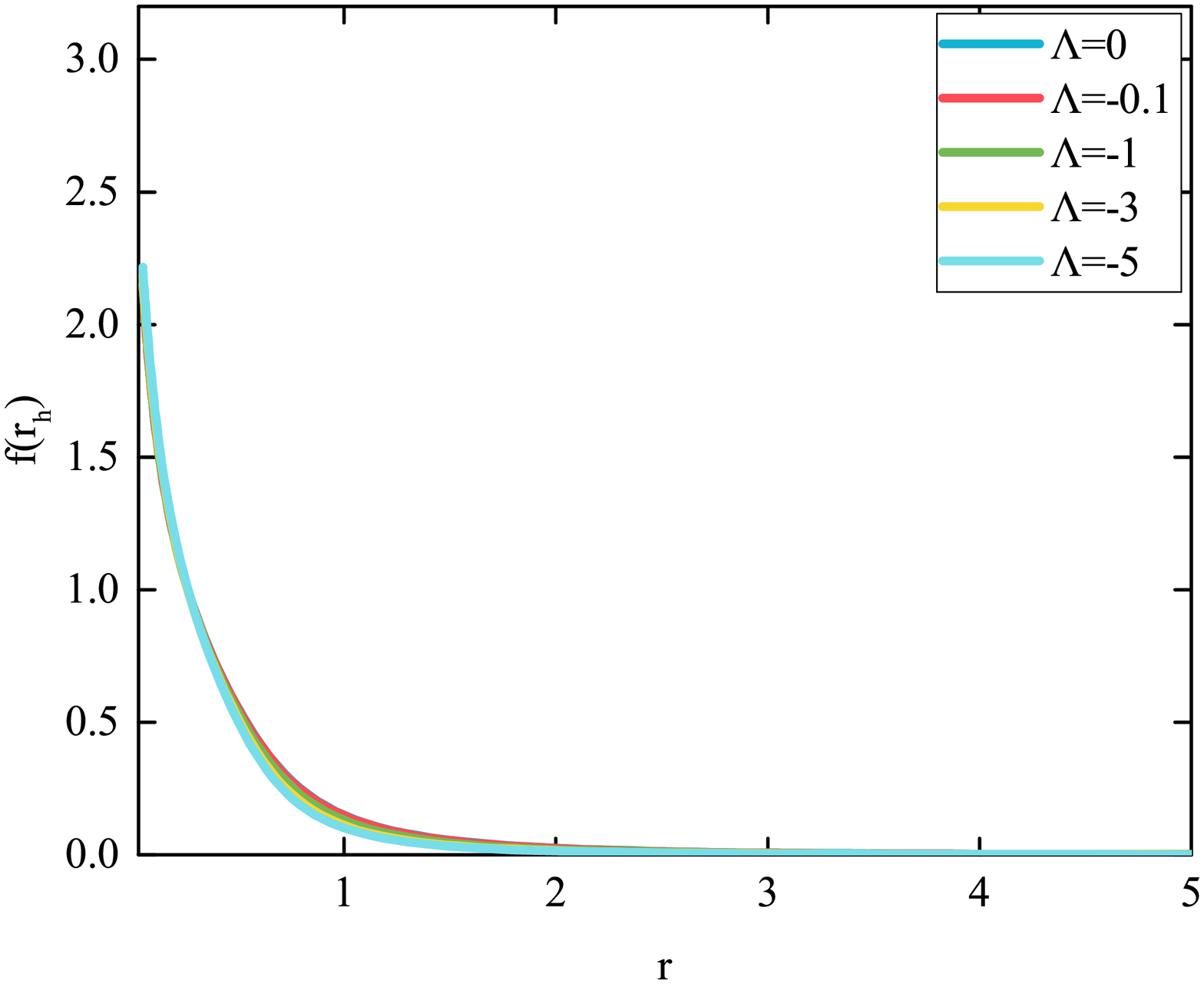}
\includegraphics[height=.26\textheight, trim = 60 20 80 50, clip = true]{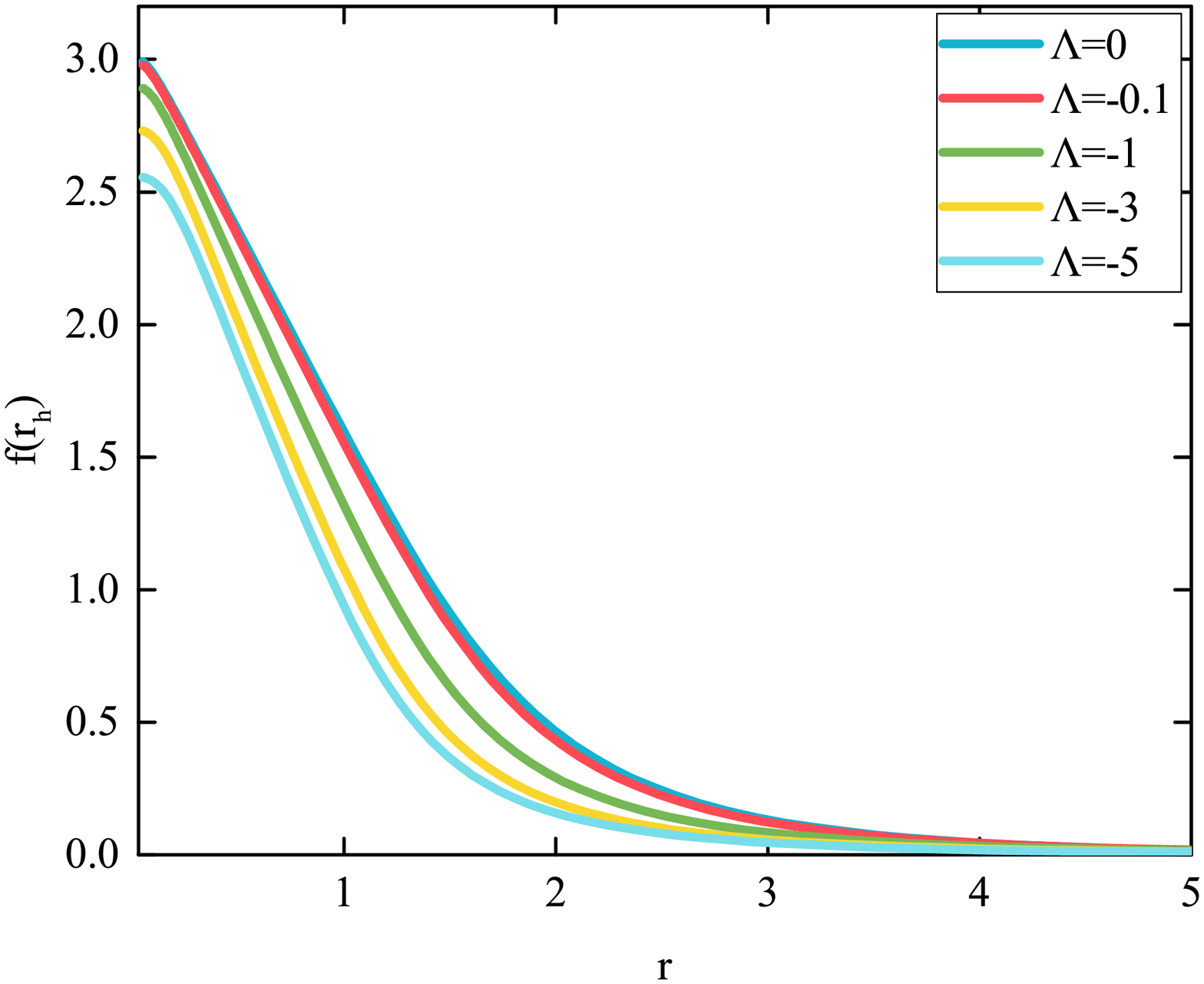}
\includegraphics[height=.26\textheight, trim = 60 20 80 50, clip = true]{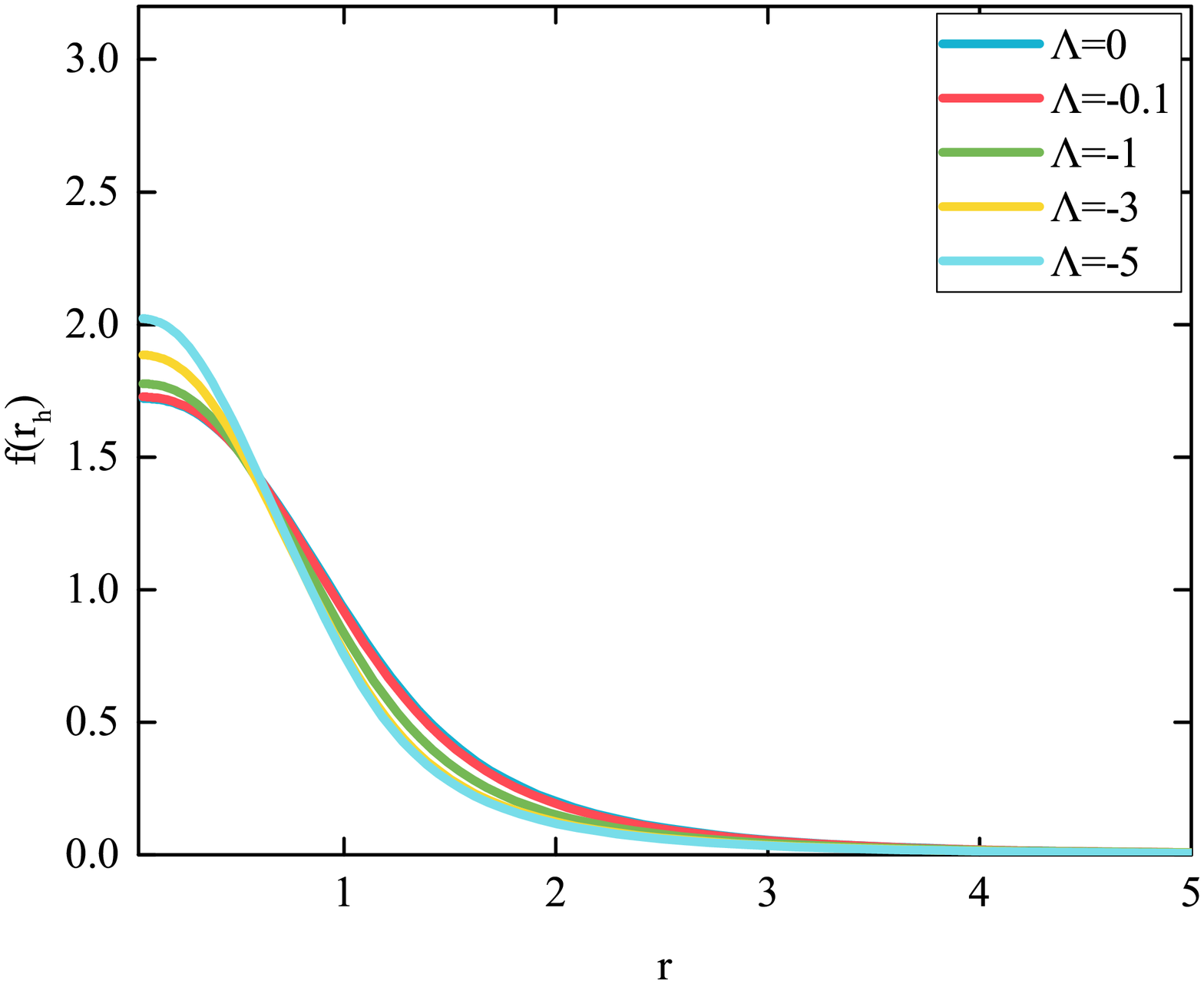}
\end{center}
\caption{\small
The Skyrme  profile functions $f(r)$ is plotted for several sets of  values of
cosmological constant $\Lambda$ at $\alpha=r_h=0.05$ and $a=b=m_\pi=1$. The left and right columns show the solutions on the
lower and upper branches solutions, respectively.
The upper row corresponds to the $c=0$ Einstein-Skyrme submodel, the lower row corresponds to the
general $c=1$ model.
}
\end{figure}

Note that for any given values of the  horizon radius $r_h$ and the effective gravitational coupling $\alpha$ there is a
maximal value of the cosmological constant $\Lambda$, for which the solutions may exist. This value monotonically decreases
with increasing of $\alpha$, however its dependency on the coupling constant $c$ is
not monotonical, it possess a maximum at some value of this parameter,
as illustrated in Fig.~\ref{lambdamax}. We also recall that the solutions may exist only for values of $\alpha$ less than
some maximal value $\alpha_{cr}$,
which decreases with increasing $r_h$. Again, with increase of $\Lambda$ the critical value of $\alpha$ is decreasing.

It might also be interesting to consider the limiting cases $a=0,~b=0$ and $a=b=0$, for which the scaling properties of the
corresponding reduced Einstein-Skyrme submodels are different. As we know, in the asymptotically flat spacetime the role of the
Skyrme term ${\mathcal L}_4$ is special, both the submodel ${\mathcal L}_0+{\mathcal L}_2+{\mathcal L}_6$ and
the BPS submodel ${\mathcal L}_0+{\mathcal L}_6$ do not support hairy black holes \cite{Adam:2016vzf,Gudnason:2016kuu}. On the other hand,
the solutions of the reduced  ${\mathcal L}_0+{\mathcal L}_4+{\mathcal L}_6$ submodel represent a new type the black holes with compact
hair.

Extending our consideration to the case of the asymptotically AdS spacetime, we can see that
both the
maximal value of event horizon radius $r_h^{cr}$ and the  ADM mass of the solutions rapidly decrease as
the parameter $b$, which multiplies the Skyrme term, is decreasing, see Fig.~\ref{fMb}. We can conclude that in the limit
$b\to 0$ there are no hairy black hole solutions for any value of the cosmological constant $\Lambda$. We note that,
for a given $b$, the increase
of the absolute value of $\Lambda$ decreases the maximal value of the horizon radius $r_h^{cr}$.

Finally,  we investigate the behavior of the solutions as the parameter $a$, which multiplies  the quadratic term
${\mathcal L}_2$, decreases. In  Fig.~\ref{fMa} we demonstrate that the
maximal value of event horizon radius $r_h^{cr}$ and ADM mass increase as $a$ is decreasing, so we may conclude that
the hairy black holes with Skyrme
hair persist in the limiting case $a=0$ for any value of $\Lambda$. However, as in the asymptotically flat spacetime, these
configurations possess compact hair. Construction of the corresponding solutions remains currently still a numerical challenge.

We note that the role of the Skyrme term remains very special for any values of $\Lambda$. Indeed, as
both parameters $a$ and $b$ decrease simultaneously, the value of $r_h^{cr}$ is also decreasing and approaches zero.
Thus, the pattern we observed for $a=b \to 0$ is similar to the case of the  $b=0$ submodel, the hairy black holes do not exist in the
models without the Skyrme term.

%%%%%%%%%%%%%%%%%%%%%%%%%%%%%%%%%%
\section{Conclusions}
%%%%%%%%%%%%%%%%%%%%%%%%%%%%%%%%%%
The main purpose of this work was to extend the discussion of the static  gravitating Skyrmions and black holes
in the generalized Einstein-Skyrme model to the case of theory with a negative cosmological constant. The effect of introduction of the
cosmological constant  can be summarise
as follows: the increase of  $|\Lambda|$ qualitatively yields the same effects as increasing
the value of the effective gravitational coupling $\alpha$, in particular the critical value of the horizon radius $r_h^{cr}$ is decreasing
as $|\Lambda|$ increase. Not entirely surprisingly, it turns out that the hairy black holes exist for any
$\Lambda$ only if the model contains the Skyrme term ${\mathcal L}_4$. Perhaps the most unusual feature of the model considered
in this work is the existence of extremal (zero-temperature) limiting black hole solutions for some set of
values of the parameters. We show that increase of $|\Lambda|$ may eliminate these singular solutions.

We have focussed on the spherically symmetric Skyrmions of topological degree one, which represent only the simplest possible type of
solutions. As a direction for future work, it would be interesting
to study the higher charge solutions, which have very geometrical
shapes \cite{Manton-book}. As a first step towards investigation of the higher degree solutions of the
generalised Einstein-Skyrme models one can consider the axially-symmetric charge two Skyrmions \cite{Sawado:2004yq},
or sphaleron solutions of the Einstein-Skyrme model \cite{Shnir:2015aba}.

Clearly, it would be interesting to investigate the spinning Kerr black holes with Skyrme hair. The regular
self-gravitating spinning Skyrmions were considered in Ref.~\cite{Ioannidou:2006nn}, however
almost nothing is known about the  stationary rotating Skrymion black holes which should emergy from
the globally regular rotating Skyrmions. One can conjecture that a new type of hair may appear for submodels
which do not have such solutions in the static limit, similar to the situation, considered in
\cite{Herdeiro:2015waa}. It might also be interesting to study how  the (non)existence of hairy Skyrmionic black
holes could be influenced by some modification of the gravity like in higher curvature models
or in dilaton-Gauss-Bonnet gravity \cite{GB}.

While in the asymptotically flat space the lower branch solutions are stable and the upper branch solutions are unstable
\cite{Bizon:1992gb,Maeda:1993ap}, the
stability analysis of the static solutions in asymptotically AdS
is not obvious. In particular, it was shown that the cosmological constant stabilizes the black hole Skyrmion in the
$c=0$ submodel \cite{Shiiki:2005aq}. Thus, we conjecture that the same would be expected of
the solutions of the generalised model. We leave this consideration for future research.

%%%%%%%%%%%%%%%%%%%%%%%%%%%%%%%%%%%%%%%%%
\section*{Acknowledgements}
%%%%%%%%%%%%%%%%%%%%%%%%%%%%%%%%%%%%%%%%%
We would like to thank C.~Adam
and A.~Wereszczynski for relevant discussions and suggestions.
I.P. is grateful to his Masha for patience, understanding and support. Y.S. gratefully
acknowledges support from the Russian Foundation for Basic Research
(Grant No. 16-52-12012) and DFG (Grant LE 838/12-2). He would like to thank Luis Ferreira, Betti Hartmann, Olga Kichakova,
Burkhard Kleihaus, Jutta Kunz and Eugen Radu for useful discussions and valuable
comments. The work of Y.S. was supported by the FAPESP (Grant No. 15/25779-6), he would like to thank the
Instituto de F\'{i}sica de S\~{a}o Carlos for its kind
hospitality.

\end{document}